\documentclass[iop,apj,tighten,twocolappendix,numberedappendix]{emulateapj}
\usepackage[breaklinks,colorlinks,citecolor=blue,linkcolor=red]{hyperref}

\usepackage{apjfonts}

\usepackage{graphicx}
\usepackage{amsmath}
\usepackage{amssymb}
\usepackage{color}
\usepackage{mathtools}
\usepackage{ulem}

\usepackage[all]{hypcap}

\newcommand\msun{\, \rm M_\odot}

\newcommand\gyr{{\, \rm Gyr}}
\newcommand\myr{{\, \rm Myr}}

\newcommand\nbh{{N_{\rm BH}}}

\newcommand{\revision}[1]{{\color{black}#1}}

\newcommand\be{\begin{equation}}
\newcommand\ee{\end{equation}}

\def\apjl{ApJL}%
\def\apjs{ApJS}%
\def\aap{A\&A}%
\begin{document}

\title{Black Hole Mergers from Star Clusters with Top-Heavy Initial Mass Functions}
\author{\href{https://orcid.org/0000-0002-9660-9085}{Newlin C. Weatherford},\altaffilmark{1,2} \href{https://orcid.org/0000-0002-7330-027X}{Giacomo Fragione},\altaffilmark{1,2} \href{https://orcid.org/0000-0002-4086-3180}{Kyle Kremer},\altaffilmark{3,4} \href{https://orcid.org/0000-0002-3680-2684}{Sourav Chatterjee},\altaffilmark{5} \href{https://orcid.org/0000-0001-9582-881X}{Claire S. Ye},\altaffilmark{1,2}  \href{https://orcid.org/0000-0003-4175-8881}{Carl L. Rodriguez},\altaffilmark{6} \href{https://orcid.org/0000-0002-7132-418X}{Frederic A. Rasio}\altaffilmark{1,2}}
 \affil{$^1$Center for Interdisciplinary Exploration \& Research in Astrophysics (CIERA), Evanston, IL 60202, USA} 
 \affil{$^2$Department of Physics \& Astronomy, Northwestern University, Evanston, IL 60202, USA}
 \affil{$^3$TAPIR, California Institute of Technology, Pasadena, CA 91125, USA}
 \affil{$^4$The Observatories of the Carnegie Institution for Science, Pasadena, CA 91101, USA}
 \affil{$^5$Tata Institute of Fundamental Research, Homi Bhabha Road, Mumbai 400005, India}
 \affil{$^6$McWilliams Center for Cosmology, Department of Physics, Carnegie Mellon University, Pittsburgh, PA 15213, USA}

\begin{abstract}
Recent observations of globular clusters (GCs) provide evidence that the stellar initial mass function (IMF) may not be universal, suggesting specifically that the IMF grows increasingly top-heavy with decreasing metallicity and increasing gas density. Non-canonical IMFs can greatly affect the evolution of GCs, mainly because the high end determines how many black holes (BHs) form. Here we compute a new set of GC models, varying the IMF within observational uncertainties. We find that GCs with top-heavy IMFs \revision{lose most of their mass} within a few Gyr through stellar winds and tidal stripping. Heating of the cluster through BH mass segregation greatly enhances this process. We show that, as they approach complete dissolution, GCs with top-heavy IMFs can evolve into `dark clusters' consisting of mostly BHs by mass. In addition to producing more BHs, GCs with top-heavy IMFs also produce many more binary BH (BBH) mergers. Even though these clusters are short-lived, mergers of ejected BBHs continue at a rate comparable to, or greater than, what is found for  long-lived GCs with canonical IMFs. Therefore these clusters, although they are no longer visible today, could still contribute significantly to the local BBH merger rate detectable by LIGO/Virgo, especially for sources with higher component masses \revision{well into the BH mass gap. We also report that one of our GC models with a top-heavy IMF produces dozens of intermediate-mass black holes (IMBHs) with masses $M>100\,{\rm M_\odot}$, including one with $M>500\,{\rm M_\odot}$. Ultimately,} additional gravitational wave observations will provide strong constraints on the stellar IMF in old GCs \revision{and the formation of IMBHs at high redshift}.
\end{abstract}


\section{Introduction}
\label{sect:intro}
Among the densest environments in the Universe, globular clusters (GCs) are ideal laboratories to investigate the importance of stellar dynamics in forming and evolving compact objects and compact binaries \citep[see, e.g.,][]{HeggieHut2003}. Frequent dynamical encounters between cluster members are fundamental in creating and explaining the existence of a number of exotic populations, such as X-ray binaries \citep[e.g.,][]{Clark1975,Verbunt1984,Giesler2018,Kremer2018a}, radio pulsars \citep[e.g.,][]{Lyne1987,Sigurdsson1995,Ivanova2008,Ye2019}, and gravitational wave (GW) sources \citep[e.g.,][]{Rodriguez2015a,Askar2017,Banerjee2017,Fragione2018b,Kremer2019b}. 

Many results of GC modeling rely on the assumption that the stellar initial mass function (IMF) has the form of a canonical \citet{Kroupa2001} IMF. Observations, however, suggest the IMF may not be universal. For example, Milky Way GCs with low central densities appear deficient in low-mass stars \citep{demarchi2007}, while GCs in the Andromeda galaxy exhibit a trend between metallicity and mass-to-light ratio that only a non-canonical, top-heavy IMF could explain \citep{haghi2017}. Ultra-compact dwarf galaxies have large dynamical mass-to-light ratios and appear to contain an overabundance of LMXB sources \citep{dabrin2009}. Our Galactic center provides an extreme example of a non-canonical IMF; stars formed there in the past few Myr have masses consistent with a top-heavy mass function \citep{bartko2010}. All these observations can be explained if the stellar IMF becomes increasingly top-heavy with decreasing metallicity and increasing gas density of the pre-GC cloud \citep{marks2012}. This is theoretically expected given the Jeans mass instability in molecular clouds \citep{larson1998} and self-regulation of accretion onto forming stars \citep{adams1996a,adams1996b}. Metallicity plays a decisive role by regulating line-emission cooling in the collapsing gas cloud and radiation pressure against stellar accretion \citep{demarchi2017}.

The IMF has long been known to strongly impact the dynamical evolution and survival of GCs. Already \citet{ChernoffWeinberg1990} found enhanced mass loss rates in cluster models with flatter IMFs, due to increased winds from high-mass stars and faster halo expansion and evaporation. More recent results show that GC models with top-heavy IMFs dissolve much faster than models with canonical IMFs \citep{chatterjee2017,giersz2019}. In particular, clusters with top-heavy IMFs produce more numerous and more massive black holes (BHs). Crucially, this promotes the BH-burning process, in which strong dynamical encounters with BHs provide energy to stellar populations, inflating the cluster halo \citep{mackey2007,mackey2008,breene2013,Kremer2019d}. Taken to an extreme by a top-heavy IMF, this mechanism will force rapid and unstable evaporation through the tidal boundary and early cluster dissolution \citep{giersz2019}.

The impact on cluster evolution of varying BH abundance has recently been further analyzed via a combination of analytical calculations and $N$-body simulations \citep{breene2013,wang2020}. However, these studies only consider idealized star clusters with two components (stars and BHs) and no stellar evolution, which plays a crucial role in cluster disruption \citep[e.g.,][]{ChernoffWeinberg1990}. Apart from the few models with non-canonical IMFs examined by \citet{chatterjee2017} and \citet{giersz2019}, there is no systematic study of the IMF's influence on BH populations generated via stellar evolution. In this Letter, we extend the grid of cluster models in the \texttt{CMC Cluster Catalog} \citep{kremer2020ApJS}, focusing on the role of a varying IMF. In particular, we study how non-canonical IMFs shape the evolution of GCs, their BH populations, and the number of dynamically produced BBH mergers.

The Letter is organized as follows. In Section~\ref{sect:models}, we describe the parameters of the numerical models we evolve. In Section~\ref{sect:evol}, we analyze the structural evolution of clusters and describe the properties of their black hole populations\revision{, including BBH mergers}. Finally, in Section~\ref{sect:conc}, we discuss the implications of our findings and lay out our conclusions.

\section{Cluster models}
\label{sect:models}

To evolve our cluster models, we use \texttt{CMC} (Cluster Monte Carlo), a H\'{e}non-type Monte Carlo code that includes the newest prescriptions for wind-driven mass loss, compact object formation, and pulsational-pair instabilities \citep[see][and references therein]{Henon1971a,Henon1971b,Joshi2000,Joshi2001,Fregeau2003,Chatterjee2010,Chatterjee2013,Pattabiraman2013,Rodriguez2015a,kremer2020ApJS}.

We generate $15$ independent cluster simulations varying the initial number of particles (single stars plus binaries) $N_{\rm i}$ and initial total mass $M_{\rm i}$, with uniform initial cluster virial radius ($r_v=1\,\rm{pc}$), metallicity ($Z=0.1\,{\rm Z}_\odot$), and Galactocentric distance\footnote{Assuming a Milky Way-like potential \citep[e.g.,][]{Dehnen1998}.} ($R_{\rm{gc}}=8\,\rm{kpc}$). These choices of $r_v$, $R_{\rm gc}$, and $Z$ are known to result in models that closely match typical Milky Way GCs when using a canonical IMF \citep{kremer2020ApJS}.

We assume that all models are initially described by a King profile with concentration $W_0 = 5$ \citep{King1962}. Primary stellar masses are sampled from the \citet{Kroupa2001} multi-component power-law IMF,
\begin{equation}
\xi(m)\propto
\begin{cases}
{m}^{-1.3}& \text{$0.08\le m/\mathrm{M}_\odot\leq 0.5$}\\
{m}^{-2.3}& \text{$0.5\le m/\mathrm{M}_\odot\leq 1.0$}\\
{m}^{-\alpha_3}& \text{$1.0\le m/\mathrm{M}_\odot\leq 150.0$}\,.
\end{cases}
\label{eqn:imf}
\end{equation}
We choose three different values for $\alpha_3=(1.6,2.3,3.0)$, corresponding to the 95\% confidence interval around the canonical value, $\alpha_3=2.3$ \citep{Kroupa2001}. We fix the primordial stellar binary fraction to $f_b=5\%$ and draw secondary masses from a uniform distribution in mass ratio \citep[e.g.,][]{DuquennoyMayor1991}. Binary orbital periods are sampled from a log-uniform distribution, with orbital separations ranging from near contact to the hard/soft boundary, while binary eccentricities are thermal \citep[e.g.,][]{Heggie1975}. \revision{For details on stellar evolution in \texttt{CMC}, see \citet{kremer2020ApJS}. We compute GW recoil kicks for BH merger products following the methods described in \citet{Rodriguez2019} and references therein. We assume all BHs have zero natal spin while BH merger products are assigned new spins of $\sim0.7$ \citep[e.g.,][]{Berti2007,Tichy2008,Lousto2010}, which are then taken into account if they merge again.}

\begin{table*}
\caption{Model initial conditions and data on cumulative BH formation and BBH mergers from $0 \leq t \leq 13\,\gyr$; upper-IMF slope $\alpha_3$, initial number of particles $N_i$, initial mass $M_i$, total BHs formed $\nbh$,  total BBH mergers $N_{\rm BBH}^{\rm (merg)}$, ejected BBH mergers $N_{\rm BBH,ejected}^{\rm (merg)}$, \revision{BBH mergers with a `mass-gap' BH component ($M_{\rm BH}>40.5\,{\rm M_\odot}$) formed via stellar collisions $N_{\rm BBH,gap,coll}^{\rm (merg)}$, multiple-generation BBH mergers with a mass-gap BH component $N_{\rm BBH,gap,2G+}^{\rm (merg)}$, and total BBH mergers at redshifts $z<1$, $N_{\rm BBH}^{\rm (merg)}(z<1)$, assuming that all GCs were born $13\,\gyr$ ago. Note that the BBH merger totals in the last 5 columns for models 1-6 are all \textit{lower limits} since we do not consider mergers resulting from dynamics in `dark clusters' (see Section~\ref{sect:rates}).}}
\centering
\begin{tabular}{lcccc|ccccc}
\hline
 & &      &   &  & \multicolumn{5}{c}{BBH mergers} \\
& $\alpha_3$ & $N_{\rm i}$ ($10^5$) & $M_{\rm i}$ ($10^5\,\msun$) & $N_{\rm BH}$ & $N_{\rm BBH}^{\rm (merg)}$ & $N_{\rm BBH,ejected}^{\rm (merg)}$ & \revision{$N_{\rm BBH,gap,coll}^{\rm (merg)}$} & \revision{$N_{\rm BBH,gap,2G+}^{\rm (merg)}$} & \revision{$N_{\rm BBH}^{\rm (merg)}(z<1)$} \\
\hline\hline
1 & 1.6 &  4    &  8.5 & 11263 &  95 &  58 &  1 &  7 &  7 \\
2 & 1.6 &  8    & 17.0 & 22539 & 222 & 132 &  7 & 26 & 19 \\
3 & 1.6 & 16    & 34.2 & 45358 & 434 & 195 & 32 & 83 & 14 \\
4 & 1.6 &  1.1  &  2.4 &  3224 &  20 &  15 &  0 &  2 &  0 \\
5 & 1.6 &  2.3  &  4.8 &  6319 &  57 &  42 &  1 &  3 &  5 \\
6 & 1.6 &  4.5  &  9.6 & 12764 & 111 &  73 &  5 & 10 & 11 \\
\hline
7 & 2.3 &  4    &  2.4 &  1114 &  38 &  17 &  1 &  3 &  5 \\
8 & 2.3 &  8    &  4.8 &  2252 &  91 &  36 &  0 &  9 & 13 \\
9 & 2.3 & 16    &  9.7 &  4518 & 233 & 106 &  1 & 23 & 42 \\
\hline
10 & 3.0 &  4   &  1.6 &   113 &   6 &   0 &  0 &  1 &  0 \\
11 & 3.0 &  8   &  3.3 &   247 &   9 &   3 &  0 &  1 &  2 \\
12 & 3.0 & 16   &  6.6 &   499 &  32 &   6 &  1 &  2 &  1 \\
13 & 3.0 &  5.9 &  2.4 &   188 &   9 &   1 &  0 &  1 &  0 \\
14 & 3.0 & 11.8 &  4.8 &   341 &  28 &   3 &  1 &  1 &  2 \\
15 & 3.0 & 23.6 &  9.7 &   744 &  71 &   9 &  2 &  5 &  7 \\
\hline
\end{tabular}
\label{tab:models}
\end{table*}

Table~\ref{tab:models} summarizes the model parameters considered in our work. Each simulation evolves to a final time $T_{\rm H}\approx14$ Gyr, unless the cluster disrupts. In these cases, which occur only for our models with $\alpha_3 = 1.6$, simulation terminates once there are fewer than 300 particles per CPU, the default particle count used to compute local densities throughout the cluster. This typically corresponds to a few$\times 10^4$ stars in the GC, beyond the point where the assumptions of the Monte Carlo method -- namely spherical symmetry and a relaxation timescale longer than the dynamical timescale -- start to break down \citep{chatterjee2017}. However, the exact cut-off point in the simulation does not affect the general evolutionary trends of the dissolution process. Further accurate evolution of the remaining `dark star cluster' would require a switch to direct $N$-body methods \citep[e.g.,][]{banerjee2011}.

\section{Results}
\label{sect:evol}

In this Section, we discuss how different IMFs affect the evolution of the cluster mass, size, and BH population. We do not find significant qualitative differences between the cases where we keep the initial number of particles constant (models 1-3, 7-9, and 10-12 in Table~\ref{tab:models}) or the initial total mass constant (models 4-9 and 13-15).

\begin{figure*} 
\centering
\includegraphics[scale=0.645]{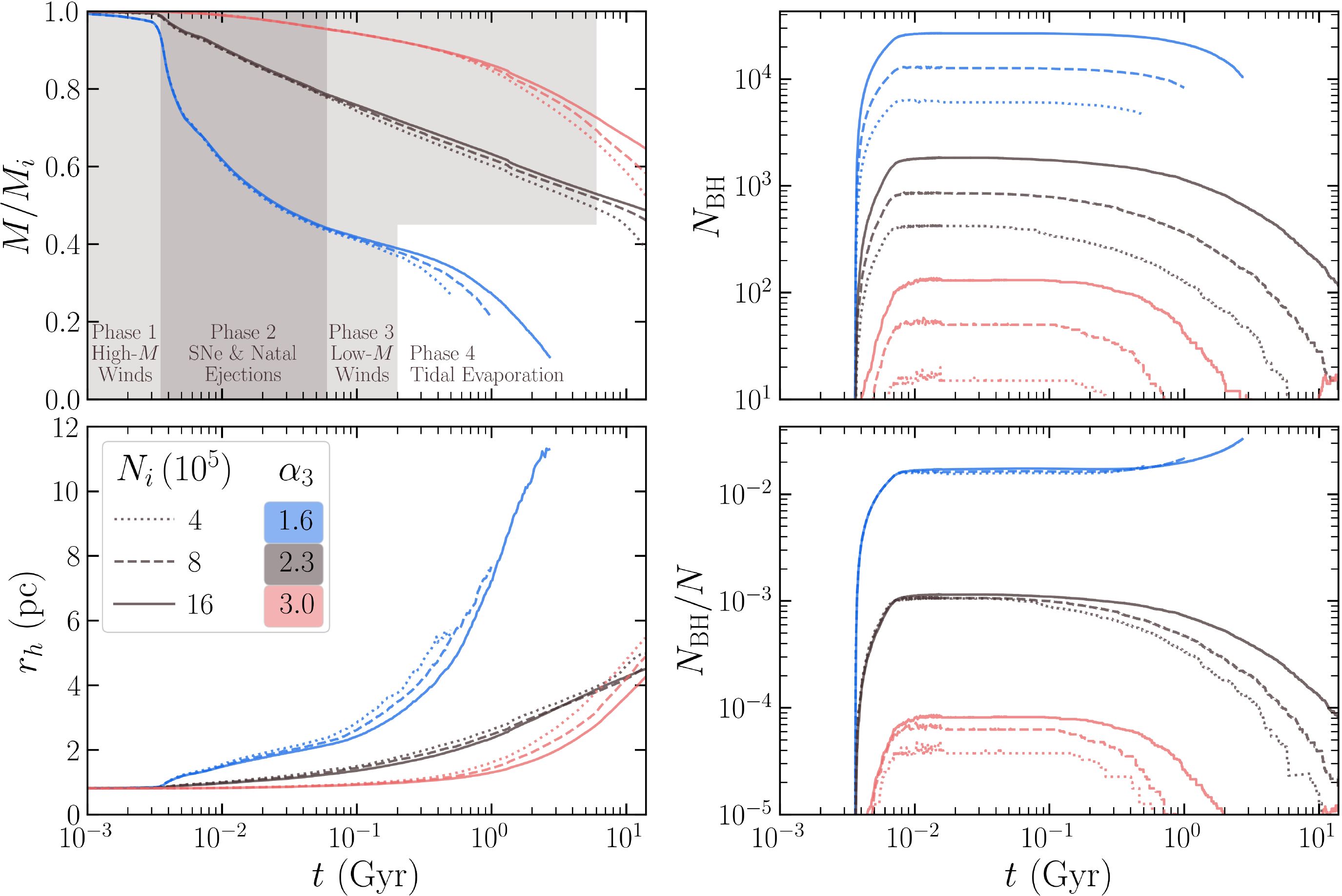}
\caption{Evolution of the total mass (top-left), half-mass radius $r_h$ (bottom-left), number of BHs $N_{\rm BH}$ (top-right), and fractional number of BHs $N_{\rm BH}/N$ (bottom-right) for cluster models with different values of $\alpha_3$ (blue $1.6$, black $2.3$, red $3.0$). Different line styles represent different initial number of particles (dotted $4\times10^5$, dashed $8\times10^5$, solid $1.6\times10^6$). The simulations terminate at $T_H=14\,\gyr$, except for $\alpha_3=1.6$, where we cut off the plot just before $r_h$ starts to drop sharply (around this point, assumptions in our Monte Carlo methods stop being valid; see text). In the top-left panel, the four mass loss phases discussed in the text are highlighted via shaded intervals and the dominant mass loss mechanisms indicated. The bifurcation of the third phase is due to decreased duration of this phase for $\alpha_3=1.6$.}
\label{fig:evol}
\end{figure*}

\subsection{Cluster evolution}

Variations in the high end of the IMF can dramatically affect cluster evolution and survival, especially by setting the number of BHs that are formed and retained, determining the degree to which BH-burning regulates the cluster's energy reservoir \citep{chatterjee2017}. In this process, BHs quickly mass-segregate into the cluster core on a sub-$\gyr$ timescale, \revision{forming a central BH population that undergoes frequent phases of core collapse and re-expansion} \citep{morscher2015}. \revision{During these events, the BHs mix with the rest of the cluster and provide} energy to passing stars in scattering interactions \citep{breene2013}. \revision{Cumulatively, these interactions inflate the cluster halo and force faster evaporation through the tidal boundary} \citep{chatterjee2017,giersz2019,wang2020}.

In Figure~\ref{fig:evol}, we show the evolution of the total mass (top-left panel) for star clusters with different $N_i$ and $\alpha_3$. Clusters evolve very differently depending on the choice of $\alpha_3$. The models with $\alpha_3=2.3$ and $3$ survive until $T_{\rm H}$, while the models with $\alpha_3=1.6$ appear to disrupt at about $t\sim 1\,\gyr$ with faster disruption occurring for smaller $N_i$. \footnote{While these models may not survive with significant mass beyond a few $\gyr$, models with larger $N_i$\revision{, larger Galactocentric distance, and/or smaller virial radius} could feasibly survive much longer.} The different fates of the cluster models are understandable given the mechanisms that power cluster mass loss at different epochs:
\begin{enumerate}
\item Initially, mass is lost primarily via stellar winds from massive stars \citep[e.g.,][]{ChernoffWeinberg1990}. This relatively quiet first phase is visible  early in the top-left panel of Figure~\ref{fig:evol}, terminating around $t \approx 3\,\myr$.
\item The second phase of mass loss, extending from $3\,\myr \lesssim t \lesssim 60\,\myr$, is driven primarily by Type II supernovae (SNe) accompanying compact object formation. This phase most clearly depends on the choice of $\alpha_3$. For $\alpha_3 = 3$, SNe-driven mass loss from few massive stars is negligible compared to ongoing stellar wind mass loss among the far more numerous low-mass stars. For $\alpha_3=1.6$, however, the SNe-driven mass loss phase is nearly catastrophic, causing the cluster to lose close to half its original mass by the time compact object formation slows ($t \approx 60\,\myr$). Ejections of lighter compact objects due to natal kicks also contribute to mass loss in this phase starting at $t\approx 6\,\myr$, causing a plateau in $\nbh$ (top-right panel).
\item A `third phase' of mass loss begins at the end of rapid compact object formation. Really just a return of the first phase, this period of mass loss is once again dominated by stellar winds, this time from lower-mass stars. For the models where $\alpha_3 = 2.3$ or $3$, stellar winds continue to account for most of the mass loss until about $6\,\gyr$. The wind-dominated mass loss phase ends much earlier for the $\alpha_3$ = 1.6 models -- around $t \approx 200\,\myr$.
\item The fourth and final phase of mass loss is driven by evaporation through the tidal boundary. This phase depends strongly on the IMF due to its influence on the number of BHs formed, and thereby the degree to which BH-burning inflates the halo. In the case of extreme BH-burning, the cluster quickly expands to fill its Roche lobe within the assumed Galactic potential and starts to tidally disrupt, as seen in the models with $\alpha_3 = 1.6$. \revision{Note that similar accelerating mass loss is also seen at late times in the models with $\alpha_3 = 3$. In these models, however, the runaway process of halo evaporation and tidal contraction is not stimulated by extreme BH-burning. Instead, these top-light, low-mass models simply start out with much smaller tidal radii. Hence they see significant evaporation rates relatively quickly compared to canonical models ($\alpha_3=2.3$), despite forming far fewer BHs.}
\end{enumerate}

The evolution of the half-mass radius $r_h$ (bottom-left panel, Figure~\ref{fig:evol}) also reflects this general picture. In all models, $r_h$ expands as a consequence of stellar mass loss. Enhanced BH-burning in models with $\alpha_3 = 1.6$ drives faster cluster expansion such that these models overflow their Roche lobes after a few hundred $\myr$. Due to more energetic BH-burning, these clusters lose about $80\%$ of their mass by $t\sim 1$ Gyr. Models with $\alpha_3=3$ do not exhibit rapid expansion, evolving more gradually as a result of reduced massive star formation and the lack of a significant central BH population.

\begin{figure} 
\centering
\includegraphics[scale=0.42]{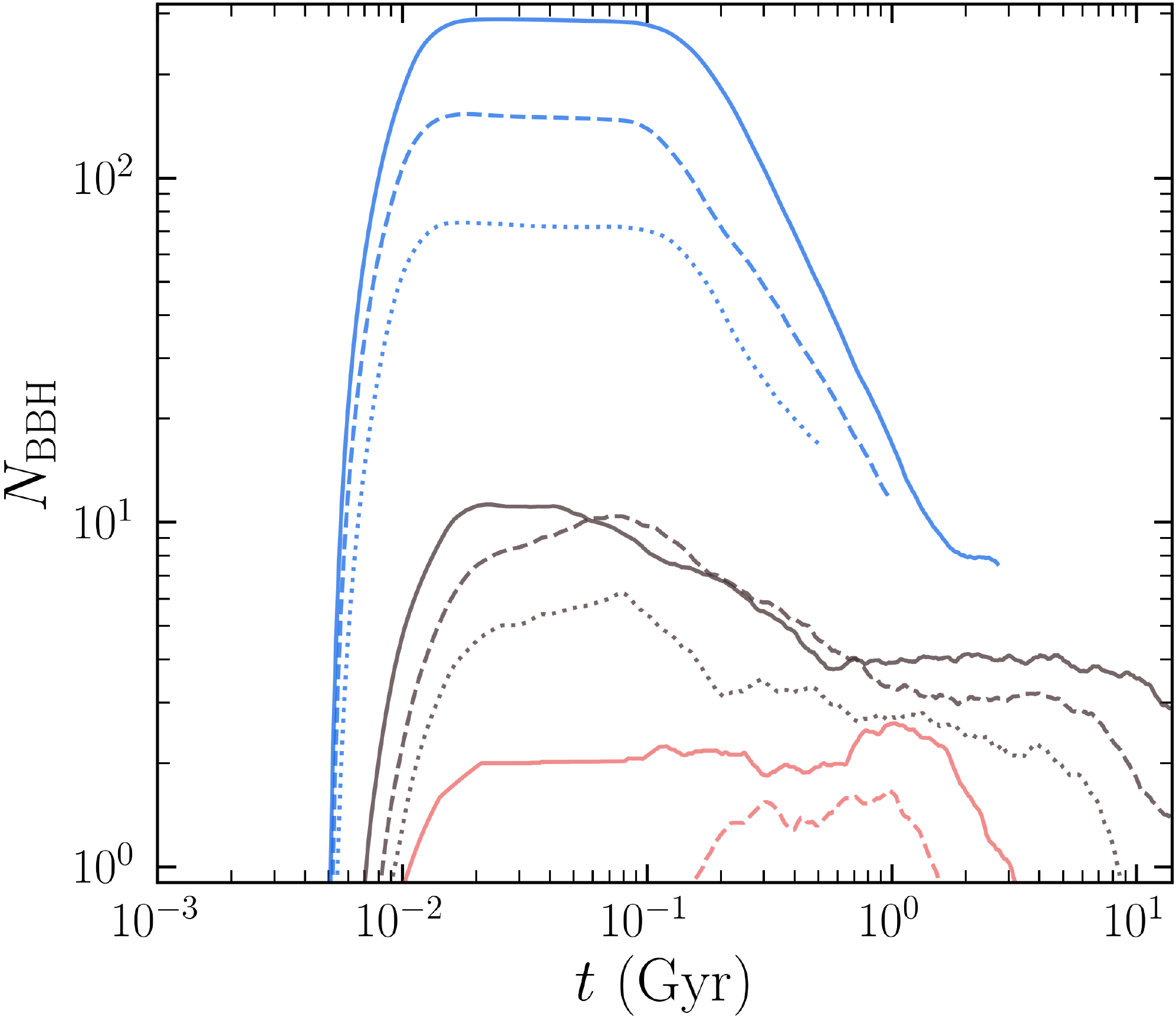}
\caption{Evolution of the number of binary black holes $N_{\rm BBH}$ bound to the star clusters with different values of $\alpha_3$ (blue $1.6$, black $2.3$, red $3.0$) and different initial number of particles ($N_i$). Different line styles represent different $N_i$, as in Figure~\ref{fig:evol}. \revision{Since BBHs continuously form and disrupt in central scattering interactions, we apply a rolling average over every $10^4$ time-steps to smooth the curves at low $N_{\rm BBH}$.}}
\label{fig:bbh}
\end{figure}

\subsection{Black hole population}

In the top-right panel of Figure~\ref{fig:evol}, we show the evolution of the number of BHs $\nbh$. Clusters with top-heavy IMFs produce more high-mass stars and thereby more BHs. As described above, natal kicks start to eject newly-formed BHs after $t\approx 6\,\myr$, causing $\nbh$ to plateau for each model. At this point, we find that $\nbh$ in models with $\alpha_3=1.6$ is $\sim 10$ times higher than in models with $\alpha_3=2.3$ and $\sim 100$ times higher than in models with $\alpha_3=3$ (see also the cumulative number of BHs formed in column 4 of Table~\ref{tab:models}). During this phase of roughly constant $\nbh$, dynamical friction causes BHs to segregate to the cluster center. This phase's duration depends on $\alpha_3$ since clusters with top-heavy IMFs have higher average stellar mass and correspondingly longer segregation timescales. Once the BHs have segregated to the core, they are gradually ejected via strong dynamical encounters.

In the bottom-right panel of Figure~\ref{fig:evol}, we also show the evolution of the fractional number of BHs $\nbh/N$. In models with $\alpha_3=2.3$ or $3$, $\nbh/N$ decreases in time as strong encounters preferentially eject massive objects (e.g., BHs) from the cluster core. In models with $\alpha_3=1.6$, however, $\nbh/N$ increases in time due to earlier tidal evaporation. By the moment these clusters retain only $\sim20\%$ of their original mass, $\nbh/N \approx 3$--$4\%$, corresponding to a significant fraction of cluster mass (up to $75\%$ by the time $r_h$ peaks). Note that the number and mass fraction of BHs at late times depends on the cluster dissolution process beyond this point, which our Monte Carlo methods are not designed to address. It is nevertheless clear that clusters with top-heavy IMFs can evolve through a `dark cluster' stage during which their total mass is dominated by BHs \citep[e.g.,][]{banerjee2011}.

Regardless of age, individual BHs are only directly detectable if they reside in binaries, either via detection of GWs from BBH mergers or through the presence of non-BH (nBH) companions. While dense star clusters are expected to be efficient factories of BBH and BH–nBH binaries, their numbers typically remain small as their dynamical assembly competes with their disruption and ejection \citep[e.g.,][]{downing2010,downing2011,morscher2015,Kremer2018a}. \revision{However, the choice of IMF greatly impacts BH binary formation.}

In Figure~\ref{fig:bbh}, we plot the evolution of the number of BBHs $N_{\rm BBH}$. Clusters with top-heavy IMFs produce more BBHs via enhanced formation of high-mass stars and BHs. \revision{$N_{\rm BBH}$ plateaus a little later than $\nbh$ (top-right panel, Figure~\ref{fig:evol}), a delay reflecting the dynamical assembly of BBHs. An equilibrium between $N_{\rm BBH}$ and $\nbh$ is apparent during this plateau, with of order one BBH for every 100 bound BHs in the cluster. Subsequently, $N_{\rm BBH}$ decreases more rapidly for clusters with higher $\nbh$, as evident by the slope past the plateau. This is unsurprising since BH-burning, enhanced for clusters with higher $\nbh$, is characterized by periodic collapse and re-expansion of the central BH population, a process that disrupts and ejects many BBHs \citep{morscher2015,chatterjee2017}.}

\begin{figure}
\centering
\includegraphics[scale=0.42]{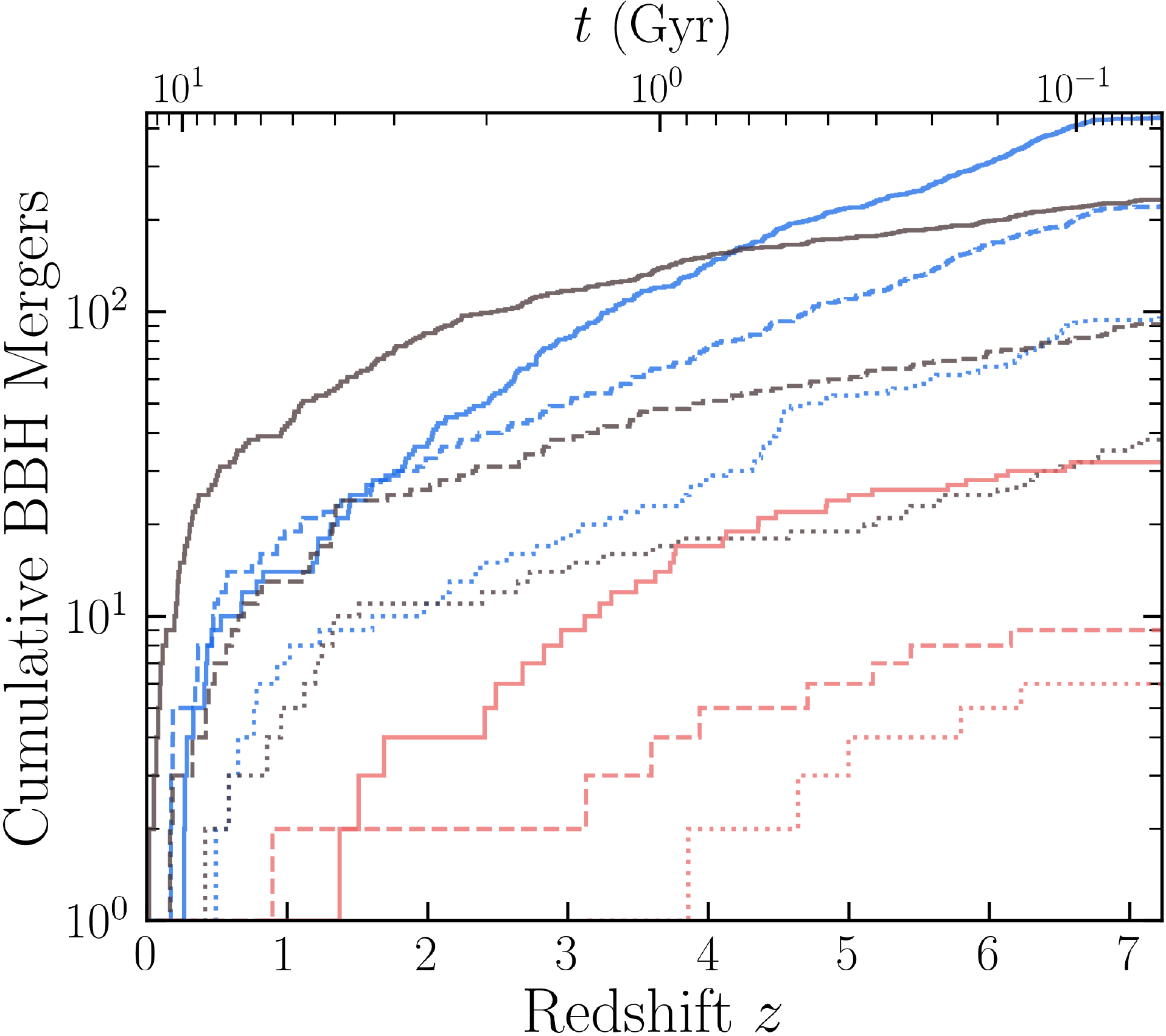}
\caption{\revision{Cumulative binary black hole mergers with respect to redshift (and time) for star clusters with different values of $\alpha_3$ (blue $1.6$, black $2.3$, red $3.0$) and different initial number of particles ($N_i$). Different line styles represent different $N_i$, as in Figure~\ref{fig:evol}. All model GCs were assumed to be born $13\,\gyr$ ago and appropriate redshifts were then computed using \texttt{Astropy}'s \revision{\citep{Astropy2013}} cosmology calculator under the flat $\Lambda$CDM model with $H_0=69.6\,{\rm km/s/Mpc}$ and $\Omega_{\rm matter} = 0.286$ \citep[see, e.g.,][]{Bennett2014}.  Values for $\alpha_3 = 1.6$ are lower limits (see Section~\ref{sect:rates}).}}
\label{fig:tinsp}
\end{figure}

\revision{\subsection{BBH merger rates} \label{sect:rates}} 
\revision{We report in Table~\ref{tab:models}, column 5, the cumulative numbers of BBH mergers $N_{\rm BBH}^{\rm (merg)}$ in each model through $13\,\gyr$. It is readily apparent that each reduction in $\alpha_3$ generally increases $N_{\rm BBH}^{\rm (merg)}$, with some nuance introduced depending on which values of $\alpha_3$ are compared and whether $N_i$ or $M_i$ is held constant in the comparison. Most notably, $N_{\rm BBH}^{\rm (merg)}$ in short-lived clusters with $\alpha_3 = 1.6$ is comparable to or greater than the number in their longer-lived cousins with $\alpha_3 = 3$. This is largely due to the sheer number of merging BBHs they eject (Table~\ref{tab:models}, column 6).

Even though clusters with top-heavy IMFs} rapidly dissolve, the large number of merging BBHs they eject suggests they could contribute significantly to the total BBH merger rate in the local Universe. \revision{Indeed, we show in Figure~\ref{fig:tinsp} that the distribution of BBH mergers across redshift $z$ is roughly comparable between the models with $\alpha_3=1.6$ and $2.3$, even for $z<1$. The number of BBH mergers in each model at $z<1$ is listed in column 9 of Table~\ref{tab:models}: a total of 40 across models 1-3 with $\alpha_3=1.6$ vs 60 across models 7-9 with $\alpha_3=2.3$.} This comparison is quite conservative \revision{since the merger counts for the models with $\alpha_3=1.6$ are \textit{lower limits}. First,} we only consider BBHs ejected prior to the simulation's end, ignoring the potentially substantial contribution to the $z<1$ merger rate from the large number of bound BHs left in the `dark phase' of these clusters' evolution. Many of these BHs could merge after ejection as binaries. Second, to compute redshift we assume the clusters were uniformly born $13\,\gyr$ ago, ignoring entirely the possibility that clusters with top-heavy IMFs may be born more recently \citep{elbadry2019}. It is therefore plausible that clusters born with top-heavy IMFs could still contribute significantly to the local BBH merger rate even if they are dissolved by the present day \citep{Fragione2018b}. 

\revision{Since the above comparisons of $z<1$ merger counts are model-dependent, it is useful to more generally estimate the BBH merger rate expected in clusters with top-heavy IMFs. We do so using the total number of BBH mergers from each model (which are still lower limits in the case of $\alpha_3 = 1.6$). The rate due to clusters with $\alpha_3=x$ at a given redshift $z$ can be approximated as $\Gamma_x (z) \approx \langle N_{\rm BBH}^{\rm (merg)} / M_i \rangle \rho_{\rm SF}(z) f_{\rm SF} f_{\rm x}$. Here, $\langle N_{\rm BBH}^{\rm (merg)} / M_i \rangle$ is the mean number of BBH mergers per initial cluster mass, $\rho_{\rm SF}$ is the cosmological density of the star formation rate, $f_{\rm SF}$ is the fraction of the star formation rate assumed to occur in star clusters, and $f_{x}$ is the fraction of clusters born with $\alpha_3\approx x$. To mitigate uncertainties in the latter three terms, we simply compute $\langle N_{\rm BBH}^{\rm (merg)} / M_i \rangle$ for each value of $\alpha_3$ and express the estimated merger rates from clusters with non-canonical IMFs as ratios with respect to the better-studied rates from clusters with canonical IMFs.}

\revision{To compute $\langle N_{\rm BBH}^{\rm (merg)} / M_i \rangle$ for each $\alpha_3$, we extract the functional dependence of $N_{\rm BBH}^{\rm (merg)}$ on $M_i$ from columns~3 and~5 of Table~\ref{tab:models}. Though the number of BBH mergers per cluster scales roughly linearly with the present-day cluster mass, at least in clusters with canonical IMFs \citep[e.g.,][]{Rodriguez2015a,kremer2020ApJS}, we
find a more general power law of the form $N_{\rm BBH}^{\rm (merg)}= a (M_i)^b$ fits the data slightly better given $\chi^2$ goodness-of-fit tests. Specifically,}

\begin{equation}
N_{\rm BBH}^{\rm (merg)}
\begin{cases}
\ge (9.5\pm 1.0)\Bigr(\frac{M_i}{10^5{\rm M_\odot}}\Bigr)^{1.09\pm 0.04} & \text{$\alpha_3=1.6$}\\
=(11.8\pm 0.5)\Bigr(\frac{M_i}{10^5{\rm M_\odot}}\Bigr)^{1.31\pm 0.02} & \text{$\alpha_3=2.3$}\\
= (1.9\pm 0.7)\Bigr(\frac{M_i}{10^5{\rm M_\odot}}\Bigr)^{1.6\pm 0.2} & \text{$\alpha_3=3.0$}\,,
\end{cases}
\label{eqn:mass_scaling}
\end{equation}

\revision{\noindent where the $1\sigma$ uncertainties on the fit parameters are computed assuming Poisson uncertainties on $N_{\rm BBH}^{\rm (merg)}$ and the $\geq$ symbol indicates a lower limit for top-heavy IMFs.

Assuming the distribution of GC birth masses takes the form $dN_{\rm cluster}/dM_i \propto M_i^{-2}$ \citep{LadaLada2003}, then}

\begin{equation}
\begin{aligned}
\biggr\langle \frac{N_{\rm BBH}^{\rm (merg)}}{M_i} \biggr\rangle
&=\ \frac{a}{(10^5{\rm M_\odot})^{b}} \frac{\int_{M_L}^{M_H} M_i^{b-3}dM_i}{\int_{M_L}^{M_H} M_i^{-2}dM_i}\\
&= \begin{cases}
\frac{a M_H \ln(M_H/M_L)}{(10^5{\rm M_\odot})^2 (M_H/M_L-1)} & \text{$b=2$}\\
\frac{a M_L^{b-1}}{(2-b)(10^5{\rm M_\odot})^{b}}\Bigr[ \frac{1-(M_L/M_H)^{2-b}}{1-M_L/M_H} \Bigr] & \text{$b\neq 2$}\,,
\end{cases}
\end{aligned}
\label{eqn:rates_integrals}
\end{equation}

\revision{\noindent where $M_L$ and $M_H$ are the lower and upper limits of the mass function for clusters capable of producing BBH mergers. Both bounds are somewhat arbitrary, but the lower bound has a greater impact on the merger rate calculation since the integrands in Equation~\ref{eqn:rates_integrals} scale inversely with $M_i$. To avoid extrapolating our fit to cluster masses that rarely produce BBHs, we set the lower bound for each $\alpha_3$ as the $M_i$ that produces an average of two stars with $M>25\,{\rm M_\odot}$, given the assumed IMF. This is roughly the minimum cluster mass needed to produce a single BBH merger since most progenitors with $M>25\,{\rm M_\odot}$ will collapse to a BH. Under this definition, $M_L/(100{\rm M_\odot}) \approx (2,9,80)$ for $\alpha_3=(1.6,2.3,3)$, respectively. Although the maximum cluster mass could vary with $\alpha_3$ in principle, observations do not well-constrain this value, so we na\"{i}vely set the upper bound for all $\alpha_3$ to be $M_H=10^7{\rm M_\odot}$ \citep{Harris2014}. With these assumptions, we find that $\langle N_{\rm BBH}^{\rm (merg)} / M_i \rangle \approx (59,40,10)/(10^6\,{\rm M_\odot})$, respectively. The rates for GCs with $\alpha_3=1.6$ and $3$, relative to the rates in canonical GCs with $\alpha_3=2.3$, are then roughly given by}

\begin{equation}
\frac{\Gamma_{1.6}}{\Gamma_{2.3}} \gtrsim 1.5\,\frac{f_{{\rm GC},1.6}}{f_{{\rm GC},2.3}}\,,\ \ \ \ \ \frac{\Gamma_{3.0}}{\Gamma_{2.3}} \approx 0.25\, \frac{f_{{\rm GC},3.0}}{f_{{\rm GC},2.3}}\,.
\label{eqn:rates_ratios}
\end{equation}

\revision{Note again that the above ratio for clusters with $\alpha_3=1.6$ is a lower limit. Both ratios are also very approximate since we neither consider interdependence of $M_i$ and $\alpha_3$ nor any cumulative dependency of these ratios on the choices of initial parameters, such as virial radius, Galactocentric distance, metallicity, and binary fraction. Furthermore, $f_x$ is notoriously uncertain and could depend on redshift, metallicity, and time of birth. Nevertheless, these estimates qualitatively suggest that if $f_{1.6}>f_{2.3}\gtrsim f_{3.0}$, then the contribution to the cosmological BBH merger rate from clusters could be a bit higher than currently estimated \citep[e.g., by][]{kremer2020ApJS}. In turn, if $f_{3.0}>f_{2.3}\gtrsim f_{1.6}$, then those rate estimates could be too high already. Overall, uncertainties in the IMF may contribute about an order of magnitude to the uncertainty on the dynamical BBH merger rate \citep[see also][]{chatterjee2017}.}

Finally, it is worth commenting briefly on black hole-neutron star (BH-NS) and binary neutron star (BNS) mergers. In clusters that retain a BH population, NSs are pushed out of the dense, BH-filled core due to mass segregation. Hence the merger likelihood is significantly smaller for BH-NS and BNS mergers than for BBH mergers in most GCs \citep{ye2020}. It is therefore unsurprising that all our models produce insignificant numbers of BH-NS and BNS mergers. While our models with $\alpha_3 = 3$ -- and therefore very few formed BHs -- do exhibit about four times more frequent binary NS mergers than the models with $\alpha_3 = 2.3$, the numbers are still far too low to account for the LIGO/Virgo estimated rates \citep{ye2020,fragban2020}.

\begin{figure} 
\centering
\includegraphics[scale=0.538]{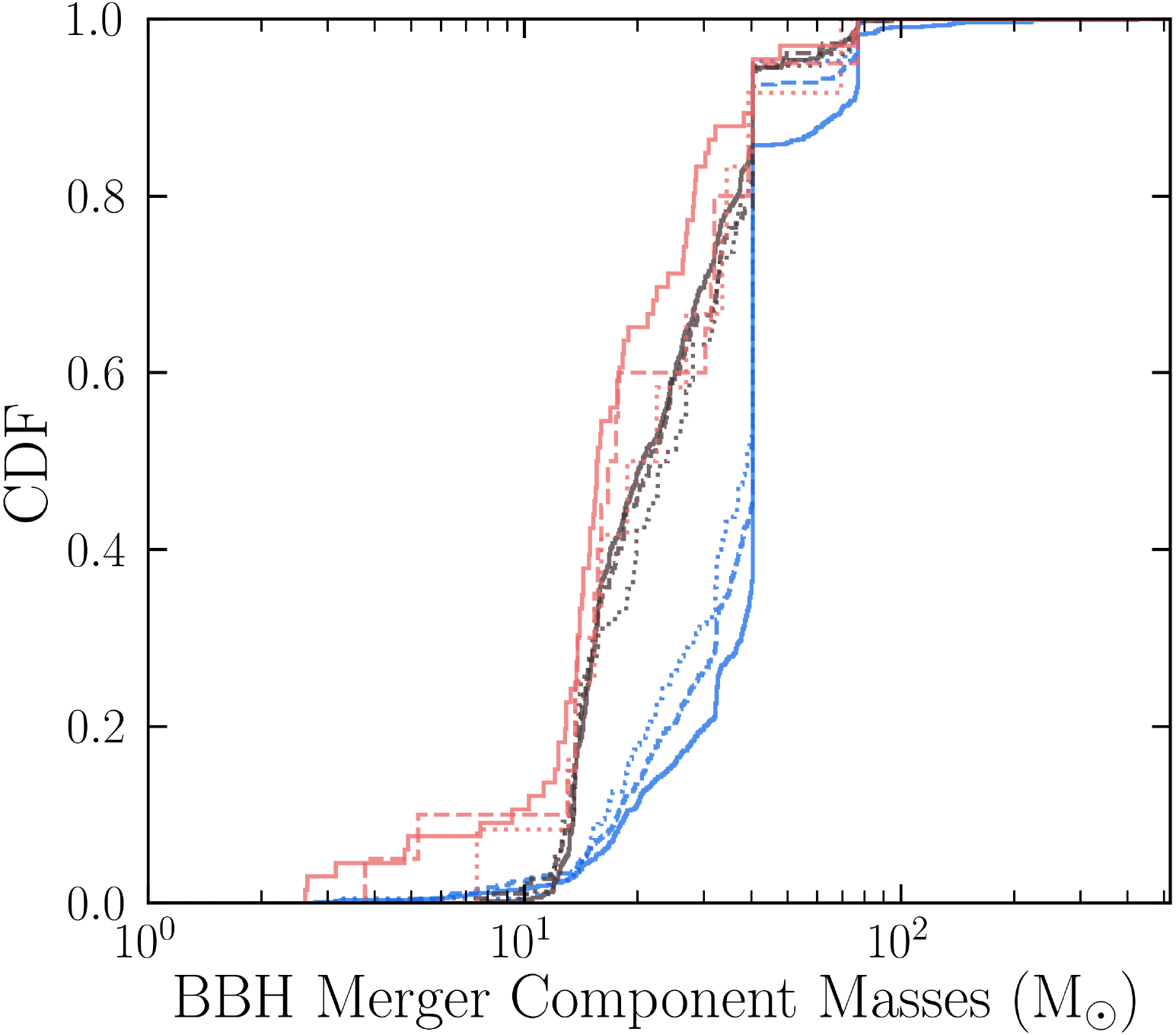}
\caption{Cumulative density functions (CDFs) of the component masses for binary black hole mergers in star clusters with different values of $\alpha_3$ (blue $1.6$, black $2.3$, red $3.0$) and different initial number of particles ($N_i$). Different line styles represent different $N_i$, as in Figure~\ref{fig:evol}.}
\label{fig:BBHmasses}
\end{figure}

\vskip 0.2truein
\revision{\subsection{BH and BBH merger masses}}
The average stellar mass is higher in clusters with top-heavy IMFs and lower in clusters with top-light IMFs. So, it is unsurprising that the average masses of BHs, BBH merger components, and BBH merger products in our models all increase with decreasing $\alpha_3$. Averaging across all our models for each $\alpha_3$ value, we find a mean BH mass at formation of $(16,12,9)\,\rm{M_\odot}$ for $\alpha_3 = (1.6,2.3,3)$, respectively. The corresponding mean masses for BBH merger components and merger products are $(38,26,21)\,\rm{M_\odot}$ and $(74,51,40)\,\rm{M_\odot}$. \revision{This trend is further exhibited in Figure~\ref{fig:BBHmasses} showing the mass distribution of BBH merger components. The vertical jump in the distributions at $M=40.5\,{\rm M_\odot}$ is due to the pile-up at the start of the `mass gap' of BHs formed from massive stars via pulsational pair-instability supernovae (PPISNe). \texttt{CMC} assumes this gap begins at $M>40.5\,{\rm M_\odot}$ \citep[for details, see][]{belczynski2016}, extending to around $120\,{\rm M_\odot}$. Hence, the first vertical jump indicates that clusters with top-heavy IMFs form significantly more BHs via PPISNe than clusters with canonical IMFs. A second, smaller jump in the distributions for top-heavy IMFs at $M \approx 77\,{\rm M_\odot}$ arises from 2nd-generation mergers with a component produced in an earlier merger between two such PPISNe-generated BHs.}

\revision{The anti-correlation between the average mass of BBH merger components and $\alpha_3$ arises for three reasons. First, clusters with top-heavy IMFs produce higher-mass stars, increasing not only the number of BHs formed, but also their average mass. Second, for initial $N$ held constant, clusters with top-heavy IMFs also have greater total mass and higher collision rates in their densely populated cores. This collision rate enhancement is obvious in Table~\ref{tab:models}, column 7, listing the number of mass gap BBH merger components formed from the stellar product of collisions \citep[e.g.,][]{dicarlo2020,krem2020coll}. While the column only lists collisionally formed merger components in the mass gap, the overall collision rate scales similarly with decreasing $\alpha_3$. Finally, the increased merger frequency in clusters with top-heavy IMFs itself increases the average mass of the components by increasing the chances that they will have already experienced a merger \citep[e.g.,][]{fraglr2020q,fraglr2020,Rodriguez2020}. 

Indeed, the cumulative number of 2nd-generation or higher (2G+) BBH mergers is also enhanced in clusters with top-heavy IMFs, as seen in Table~\ref{tab:models}, column 8, listing the number of 2G+ mergers with a component in the mass gap. For clusters with top-heavy IMFs, this is identical to the total number of 2G+ BBH mergers, though 2G+ mergers with neither component in the mass gap are common for higher $\alpha_3$. Overall, 2G+ merger totals roughly double with each decrease in $\alpha_3$ from $3$ to $2.3$ to $1.6$, while the numbers of 2G+ mergers with a mass gap component roughly quadruple.}

Though we find no trend between initial $N$ and merger component or product mass for $\alpha_3 = 2.3$ or $3$, there is a clear correlation for $\alpha_3=1.6$. In this case, for $N_i=(4,8,16)\times 10^5$, the average BBH merger product mass is $(54,62,79)\,\rm{M_\odot}$, respectively. This rise is in part due to 2G+ mergers, which account for $(7\%,12\%,18\%)$ of the BBH mergers in these respective models. In general, we find that clusters with top-heavy IMFs and high initial $N$ are especially good at efficiently producing \revision{many 2G+ mergers and many BHs in the mass gap.

The increased collision and 2G+ merger rates in clusters with top-heavy IMFs can also result in the formation of `intermediate-mass' BHs (IMBHs), which we define as BHs with masses exceeding $100\,{\rm M_\odot}$. In particular, our highest-$N$ model with $\alpha_3=1.6$ (model 3 in  Table~\ref{tab:models}) produce 46 IMBHs, including one with mass $M=537\,{\rm M_\odot}$. This particular IMBH formed in the merger of a $122\,{\rm M_\odot}$ IMBH with a $426\,{\rm M_\odot}$ IMBH, which itself formed in the merger of two IMBHs with $M\approx200\,{\rm M_\odot}$. While we intend to explore IMBH-formation in GCs with top-heavy IMFs more thoroughly in future work, we for now direct the reader's attention to our collaboration's recent study on IMBH formation in GCs \citep{Gonzalez2020}.}

\vskip 0.3truein
\section{Discussion and Conclusions}
\label{sect:conc}

Observations provide evidence that the stellar IMF may not be universal. A non-canonical IMF can greatly affect a star cluster's dynamical evolution, especially since its high end determines how many BHs form within, regulating the cluster's energy budget and dynamical clock \citep{breene2013,chatterjee2017,Kremer2019d,wang2020}. In this Letter, we have extended the \texttt{CMC Cluster Catalog} \citep{kremer2020ApJS} to examine how a varying IMF affects the evolution of star clusters and their BH populations.

We have shown that massive star clusters with top-heavy IMFs (low $\alpha_3$) are likely to lose most of their mass within a few Gyr, assuming they have low-to-average mass and Galactocentric distance for typical Milky Way GCs. The rapid mass loss during dissolution occurs in stages, first driven by stellar winds and dynamical ejections, then by evaporation of the halo through the tidal boundary. Extensive BH-burning enhances the latter stage in clusters with top-heavy IMFs, which produce many BHs. Such clusters evolve through a point where they consist mostly of BHs by mass (and up to at least $3\%$ by number).\footnote{Just like the fictional Maw cluster \citep{Anderson1994}, this `dark cluster' stage of a GC born with a top-heavy IMF may be short-lived due to the rapid pace of tidal evaporation.} Further study with direct $N$-body methods \citep[e.g.,][]{banerjee2011} is required to fully understand the evolution of these clusters. \revision{Initializing direct $N$-body simulations with the pre-dissolution states of \texttt{CMC} models could be especially useful in such an evaluation across the full cluster mass distribution.}

We note that the processes described above are also affected by the choices of initial cluster metallicity and natal kick distribution \citep{chatterjee2017}. BHs in metal-rich clusters have lower mass and do not inject as much energy into the BH-burning process as BHs in metal-poor clusters. Thus, metal-rich clusters typically have higher densities and dispersion velocities, therefore processing the BH population on shorter timescales and disrupting more binaries. Meanwhile, high natal kicks will eject most BHs from the cluster during formation. In such a case, BHs and BBHs are expected to be small in number regardless of the IMF. \revision{Top-heavy clusters with higher natal kicks also live longer; while they experience less tidal mass loss driven by BH-burning, these clusters lose even more mass due to kick-driven dynamical ejections of BHs \citep{chatterjee2017}. For a recent study of top-heavy cluster evolution featuring different natal kick assumptions, see \citet[][published during review of this paper]{Haghi2020}, who examine lower-mass clusters with direct $N-$body simulations incorporating gas expulsion physics. Notably, they find that top-heavy clusters, albeit drastically reduced in mass, may well survive to the present day if born with masses above $M_i\gtrsim 7\times10^5\,{\rm M_\odot}$. For times not too close to disruption, where \texttt{CMC}'s assumptions start to be challenged (see Section~\ref{sect:models}), our results are encouragingly compatible with those of \citet{Haghi2020}.}

Regardless of their long-term evolution and stability, we have also shown that clusters with top-heavy IMFs -- and correspondingly high BH production -- may contribute significantly to the present-day binary BH merger rate. Even though these clusters rapidly lose most of their mass within a few Gyr, mergers from ejected BBHs continue to contribute at later times \citep{Fragione2018b}, at rates comparable to or greater than those for clusters with canonical Kroupa IMFs. The rate of 2nd-generation mergers with component masses in the mass gap may be especially enhanced in top-heavy GCs, motivating the existence of more GW190521-like mergers \citep{GW190521}. \revision{In addition, the enhancement of collision rates and multiple-generation mergers in top-heavy GCs may also lead to the formation of IMBHs and even IMBH-IMBH mergers, as demonstrated in one of our models.}

\revision{In general, we have shown that the high-mass slope of the cluster birth IMF may significantly impact the exact contribution to the cosmological BBH merger rate due to cluster dynamics. Specifically, if a large fraction of clusters were born with top-heavy IMFs, the cluster-dynamics merger rate may be somewhat enhanced relative to recent estimates \citep[e.g.,][]{kremer2020ApJS}. In turn, if a large fraction of clusters were born with top-light IMFs, the cluster-dynamics merger rate may be significantly reduced.} With future observations of gravitational waves providing unique information on the BBH merger rate \citep{lvc2020cat}, it may be possible to leverage this understanding to better constrain the IMFs of old star clusters.

\section*{Acknowledgements}

This work was supported by NSF grant AST-1716762 and through the computational resources and staff contributions provided for the Quest high-performance computing facility at Northwestern University. NW acknowledges support from the CIERA Riedel Family Graduate Fellowship as well as the NSF GK-12 Fellowship Program under Grant DGE-0948017. GF acknowledges support from a CIERA Fellowship. KK is supported by an NSF Astronomy and Astrophysics Postdoctoral Fellowship under award AST-2001751. SC acknowledges  support  from  the  Department  of  Atomic Energy,  Government  of  India,  under  Project  No.~12-R\&D-TFR-5.02-0200. 

\bibliography{Weatherford21}

\begin{thebibliography}{}
\providecommand\natexlab[1]{#1}
\providecommand\JournalTitle[1]{#1}

\bibitem[{{Abbott} {et~al.}(2020{\natexlab{a}}){Abbott}, {Abbott}, {Abraham},
  {Acernese}, {Ackley}, {Adams}, {Adhikari}, {Adya}, {Affeldt}, {Agathos},
  {Agatsuma}, {Aggarwal}, {Aguiar}, {Aich}, {Aiello}, {Ain}, {Ajith}, {Akcay},
  {Allen}, {Allocca}, {Altin}, {Amato}, {Anand}, {Ananyeva}, {Anderson},
  {Anderson}, {Angelova}, {Ansoldi}, {Antier}, {Appert}, {Arai}, {Araya},
  {Areeda}, {Ar{\`e}ne}, {Arnaud}, {Aronson}, {Arun}, {Asali}, {Ascenzi},
  {Ashton}, {Aston}, {Astone}, {Aubin}, {Aufmuth}, {AultONeal}, {Austin},
  {Avendano}, {Babak}, {Bacon}, {Badaracco}, {Bader}, {Bae}, {Baer}, {Baird},
  {Baldaccini}, {Ballardin}, {Ballmer}, {Bals}, {Balsamo}, {Baltus},
  {Banagiri}, {Bankar}, {Bankar}, {Barayoga}, {Barbieri}, {Barish}, {Barker},
  {Barkett}, {Barneo}, {Barone}, {Barr}, {Barsotti}, {Barsuglia}, {Barta},
  {Bartlett}, {Bartos}, {Bassiri}, {Basti}, {Bawaj}, {Bayley}, {Bazzan},
  {B{\'e}csy}, {Bejger}, {Belahcene}, {Bell}, {Beniwal}, {Benjamin}, {Bentley},
  {Bergamin}, {Berger}, {Bergmann}, {Bernuzzi}, {Berry}, {Bersanetti},
  {Bertolini}, {Betzwieser}, {Bhand are}, {Bhandari}, {Bidler}, {Biggs},
  {Bilenko}, {Billingsley}, {Birney}, {Birnholtz}, {Biscans}, {Bischi},
  {Biscoveanu}, {Bisht}, {Bissenbayeva}, {Bitossi}, {Bizouard}, {Blackburn},
  {Blackman}, {Blair}, {Blair}, {Blair}, {Bobba}, {Bode}, {Boer}, {Boetzel},
  {Bogaert}, {Bondu}, {Bonilla}, {Bonnand}, {Booker}, {Boom}, {Bork}, {Boschi},
  {Bose}, {Bossilkov}, {Bosveld}, {Bouffanais}, {Bozzi}, {Bradaschia}, {Brady},
  {Bramley}, {Branchesi}, {Brau}, {Breschi}, {Briant}, {Briggs}, {Brighenti},
  {Brillet}, {Brinkmann}, {Brockill}, {Brooks}, {Brooks}, {Brown}, {Brunett},
  {Bruno}, {Bruntz}, {Buikema}, {Bulik}, {Bulten}, {Buonanno}, {Buscicchio},
  {Buskulic}, {Byer}, {Cabero}, {Cadonati}, {Cagnoli}, {Cahillane},
  {Calder{\'o}n Bustillo}, {Callaghan}, {Callister}, {Calloni}, {Camp},
  {Canepa}, {Cannon}, {Cao}, {Cao}, {Carapella}, {Carbognani}, {Caride},
  {Carney}, {Carullo}, {Casanueva Diaz}, {Casentini}, {Casta{\~n}eda},
  {Caudill}, {Cavagli{\`a}}, {Cavalier}, {Cavalieri}, {Cella},
  {Cerd{\'a}-Dur{\'a}n}, {Cesarini}, {Chaibi}, {Chakravarti}, {Chan}, {Chan},
  {Chandra}, {Chao}, {Charlton}, {Chase}, {Chassande-Mottin}, {Chatterjee},
  {Chaturvedi}, {Chatziioannou}, {Chen}, {Chen}, {Chen}, {Cheng}, {Cheong},
  {Chia}, {Chiadini}, {Chierici}, {Chincarini}, {Chiummo}, {Cho}, {Cho}, {Cho},
  {Christensen}, {Chu}, {Chua}, {Chung}, {Chung}, {Ciani}, {Ciecielag},
  {Cie{\'s}lar}, {Ciobanu}, {Ciolfi}, {Cipriano}, {Cirone}, {Clara}, {Clark},
  {Clearwater}, {Clesse}, {Cleva}, {Coccia}, {Cohadon}, {Cohen}, {Colleoni},
  {Collette}, {Collins}, {Colpi}, {Constancio}, {Conti}, {Cooper}, {Corban},
  {Corbitt}, {Cordero-Carri{\'o}n}, {Corezzi}, {Corley}, {Cornish}, {Corre},
  {Corsi}, {Cortese}, {Costa}, {Cotesta}, {Coughlin}, {Coughlin}, {Coulon},
  {Countryman}, {Couvares}, {Covas}, {Coward}, {Cowart}, {Coyne}, {Coyne},
  {Creighton}, {Creighton}, {Cripe}, {Croquette}, {Crowder}, {Cudell},
  {Cullen}, {Cumming}, {Cummings}, {Cunningham}, {Cuoco}, {Curylo}, {Canton},
  {D{\'a}lya}, {Dana}, {Daneshgaran-Bajastani}, {D'Angelo}, {Danilishin},
  {D'Antonio}, {Danzmann}, {Darsow-Fromm}, {Dasgupta}, {Datrier}, {Dattilo},
  {Dave}, {Davier}, {Davies}, {Davis}, {Daw}, {DeBra}, {Deenadayalan},
  {Degallaix}, {De Laurentis}, {Del{\'e}glise}, {Delfavero}, {De Lillo}, {Del
  Pozzo}, {DeMarchi}, {D'Emilio}, {Demos}, {Dent}, {De Pietri}, {De Rosa}, {De
  Rossi}, {DeSalvo}, {de Varona}, {Dhurand har}, {D{\'\i}az}, {Diaz-Ortiz},
  {Dietrich}, {Di Fiore}, {Di Fronzo}, {Di Giorgio}, {Di Giovanni}, {Di
  Giovanni}, {Di Girolamo}, {Di Lieto}, {Ding}, {Di Pace}, {Di Palma}, {Di
  Renzo}, {Divakarla}, {Dmitriev}, {Doctor}, {Donovan}, {Dooley}, {Doravari},
  {Dorrington}, {Downes}, {Drago}, {Driggers}, {Du}, {Ducoin}, {Dupej},
  {Durante}, {D'Urso}, {Dwyer}, {Easter}, {Eddolls}, {Edelman}, {Edo}, {Edy},
  {Effler}, {Ehrens}, {Eichholz}, {Eikenberry}, {Eisenmann}, {Eisenstein},
  {Ejlli}, {Errico}, {Essick}, {Estelles}, {Estevez}, {Etienne}, {Etzel},
  {Evans}, {Evans}, {Ewing}, {Fafone}, {Fairhurst}, {Fan}, {Farinon}, {Farr},
  {Farr}, {Fauchon-Jones}, {Favata}, {Fays}, {Fazio}, {Feicht}, {Fejer},
  {Feng}, {Fenyvesi}, {Ferguson}, {Fernandez-Galiana}, {Ferrante}, {Ferreira},
  {Ferreira}, {Fidecaro}, {Fiori}, {Fiorucci}, {Fishbach}, {Fisher},
  {Fittipaldi}, {Fitz-Axen}, {Fiumara}, {Flaminio}, {Floden}, {Flynn}, {Fong},
  {Font}, {Forsyth}, {Fournier}, {Frasca}, {Frasconi}, {Frei}, {Freise},
  {Frey}, {Frey}, {Fritschel}, {Frolov}, {Fronz{\`e}}, {Fulda}, {Fyffe},
  {Gabbard}, {Gadre}, {Gaebel}, {Gair}, {Galaudage}, {Ganapathy}, {Ganguly},
  {Gaonkar}, {Garc{\'\i}a-Quir{\'o}s}, {Garufi}, {Gateley}, {Gaudio},
  {Gayathri}, {Gemme}, {Genin}, {Gennai}, {George}, {George}, {Gergely},
  {Ghonge}, {Ghosh}, {Ghosh}, {Ghosh}, {Giacomazzo}, {Giaime}, {Giardina},
  {Gibson}, {Gier}, {Gill}, {Glanzer}, {Gniesmer}, {Godwin}, {Goetz}, {Goetz},
  {Gohlke}, {Goncharov}, {Gonz{\'a}lez}, {Gopakumar}, {Gossan}, {Gosselin},
  {Gouaty}, {Grace}, {Grado}, {Granata}, {Grant}, {Gras}, {Grassia}, {Gray},
  {Gray}, {Greco}, {Green}, {Green}, {Gretarsson}, {Griggs}, {Grignani},
  {Grimaldi}, {Grimm}, {Grote}, {Grunewald}, {Gruning}, {Guidi}, {Guimaraes},
  {Guix{\'e}}, {Gulati}, {Guo}, {Gupta}, {Gupta}, {Gupta}, {Gustafson},
  {Gustafson}, {Haegel}, {Halim}, {Hall}, {Hamilton}, {Hammond}, {Haney},
  {Hanke}, {Hanks}, {Hanna}, {Hannam}, {Hannuksela}, {Hansen}, {Hanson},
  {Harder}, {Hardwick}, {Haris}, {Harms}, {Harry}, {Harry}, {Hasskew},
  {Haster}, {Haughian}, {Hayes}, {Healy}, {Heidmann}, {Heintze}, {Heinze},
  {Heitmann}, {Hellman}, {Hello}, {Hemming}, {Hendry}, {Heng}, {Hennes},
  {Hennig}, {Heurs}, {Hild}, {Hinderer}, {Hoback}, {Hochheim}, {Hofgard},
  {Hofman}, {Holgado}, {Holland}, {Holt}, {Holz}, {Hopkins}, {Horst}, {Hough},
  {Howell}, {Hoy}, {Huang}, {H{\"u}bner}, {Huerta}, {Huet}, {Hughey}, {Hui},
  {Husa}, {Huttner}, {Huxford}, {Huynh-Dinh}, {Idzkowski}, {Iess}, {Inchauspe},
  {Ingram}, {Intini}, {Isac}, {Isi}, {Iyer}, {Jacqmin}, {Jadhav}, {Jadhav},
  {James}, {Jani}, {Janthalur}, {Jaranowski}, {Jariwala}, {Jaume}, {Jenkins},
  {Jiang}, {Johns}, {Johnson-McDaniel}, {Jones}, {Jones}, {Jones}, {Jones},
  {Jones}, {Jonker}, {Ju}, {Junker}, {Kalaghatgi}, {Kalogera}, {Kamai},
  {Kandhasamy}, {Kang}, {Kanner}, {Kapadia}, {Karki}, {Kashyap}, {Kasprzack},
  {Kastaun}, {Katsanevas}, {Katsavounidis}, {Katzman}, {Kaufer}, {Kawabe},
  {K{\'e}f{\'e}lian}, {Keitel}, {Keivani}, {Kennedy}, {Key}, {Khadka},
  {Khalili}, {Khan}, {Khan}, {Khan}, {Khazanov}, {Khetan}, {Khursheed},
  {Kijbunchoo}, {Kim}, {Kim}, {Kim}, {Kim}, {Kim}, {Kim}, {Kim}, {Kimball},
  {King}, {Kinley-Hanlon}, {Kirchhoff}, {Kissel}, {Kleybolte}, {Klimenko},
  {Knowles}, {Knyazev}, {Koch}, {Koehlenbeck}, {Koekoek}, {Koley},
  {Kondrashov}, {Kontos}, {Koper}, {Korobko}, {Korth}, {Kovalam}, {Kozak},
  {Kringel}, {Krishnendu}, {Kr{\'o}lak}, {Krupinski}, {Kuehn}, {Kumar},
  {Kumar}, {Kumar}, {Kumar}, {Kumar}, {Kuo}, {Kutynia}, {Lackey}, {Laghi},
  {Lalande}, {Lam}, {Lamberts}, {Landry}, {Lane}, {Lang}, {Lange}, {Lantz},
  {Lanza}, {La Rosa}, {Lartaux-Vollard}, {Lasky}, {Laxen}, {Lazzarini},
  {Lazzaro}, {Leaci}, {Leavey}, {Lecoeuche}, {Lee}, {Lee}, {Lee}, {Lee}, {Lee},
  {Lehmann}, {Leroy}, {Letendre}, {Levin}, {Li}, {Li}, {li}, {Li}, {Li},
  {Linde}, {Linker}, {Linley}, {Littenberg}, {Liu}, {Liu},
  {Llorens-Monteagudo}, {Lo}, {Lockwood}, {London}, {Longo}, {Lorenzini},
  {Loriette}, {Lormand}, {Losurdo}, {Lough}, {Lousto}, {Lovelace}, {L{\"u}ck},
  {Lumaca}, {Lundgren}, {Ma}, {Macas}, {Macfoy}, {MacInnis}, {Macleod},
  {MacMillan}, {Macquet}, {Maga{\~n}a Hernandez}, {Maga{\~n}a-Sandoval},
  {Magee}, {Majorana}, {Maksimovic}, {Malik}, {Man}, {Mandic}, {Mangano},
  {Mansell}, {Manske}, {Mantovani}, {Mapelli}, {Marchesoni}, {Marion},
  {M{\'a}rka}, {M{\'a}rka}, {Markakis}, {Markosyan}, {Markowitz}, {Maros},
  {Marquina}, {Marsat}, {Martelli}, {Martin}, {Martin}, {Martinez}, {Martynov},
  {Masalehdan}, {Mason}, {Massera}, {Masserot}, {Massinger}, {Masso-Reid},
  {Mastrogiovanni}, {Matas}, {Matichard}, {Mavalvala}, {Maynard}, {McCann},
  {McCarthy}, {McClelland}, {McCormick}, {McCuller}, {McGuire}, {McIsaac},
  {McIver}, {McManus}, {McRae}, {McWilliams}, {Meacher}, {Meadors}, {Mehmet},
  {Mehta}, {Mejuto Villa}, {Melatos}, {Mendell}, {Mercer}, {Mereni}, {Merfeld},
  {Merilh}, {Merritt}, {Merzougui}, {Meshkov}, {Messenger}, {Messick},
  {Metzdorff}, {Meyers}, {Meylahn}, {Mhaske}, {Miani}, {Miao}, {Michaloliakos},
  {Michel}, {Middleton}, {Milano}, {Miller}, {Millhouse}, {Mills}, {Milotti},
  {Milovich-Goff}, {Minazzoli}, {Minenkov}, {Mishkin}, {Mishra}, {Mistry},
  {Mitra}, {Mitrofanov}, {Mitselmakher}, {Mittleman}, {Mo}, {Mogushi},
  {Mohapatra}, {Mohite}, {Molina-Ruiz}, {Mondin}, {Montani}, {Moore}, {Moraru},
  {Morawski}, {Moreno}, {Morisaki}, {Mours}, {Mow-Lowry}, {Mozzon},
  {Muciaccia}, {Mukherjee}, {Mukherjee}, {Mukherjee}, {Mukherjee}, {Mukund},
  {Mullavey}, {Munch}, {Mu{\~n}iz}, {Murray}, {Nagar}, {Nardecchia},
  {Naticchioni}, {Nayak}, {Neil}, {Neilson}, {Nelemans}, {Nelson}, {Nery},
  {Neunzert}, {Ng}, {Ng}, {Nguyen}, {Nguyen}, {Nichols}, {Nichols}, {Nissanke},
  {Nitz}, {Nocera}, {Noh}, {North}, {Nothard}, {Nuttall}, {Oberling},
  {O'Brien}, {Oganesyan}, {Ogin}, {Oh}, {Oh}, {Ohme}, {Ohta}, {Okada},
  {Oliver}, {Olivetto}, {Oppermann}, {Oram}, {O'Reilly}, {Ormiston}, {Ortega},
  {O'Shaughnessy}, {Ossokine}, {Osthelder}, {Ottaway}, {Overmier}, {Owen},
  {Pace}, {Pagano}, {Page}, {Pagliaroli}, {Pai}, {Pai}, {Palamos}, {Palashov},
  {Palomba}, {Pan}, {Panda}, {Pang}, {Pankow}, {Pannarale}, {Pant}, {Paoletti},
  {Paoli}, {Parida}, {Parker}, {Pascucci}, {Pasqualetti}, {Passaquieti},
  {Passuello}, {Patricelli}, {Payne}, {Pearlstone}, {Pechsiri}, {Pedersen},
  {Pedraza}, {Pele}, {Penn}, {Perego}, {Perez}, {P{\'e}rigois}, {Perreca},
  {Perri{\`e}s}, {Petermann}, {Pfeiffer}, {Phelps}, {Phukon}, {Piccinni},
  {Pichot}, {Piendibene}, {Piergiovanni}, {Pierro}, {Pillant}, {Pinard},
  {Pinto}, {Piotrzkowski}, {Pirello}, {Pitkin}, {Plastino}, {Poggiani}, {Pong},
  {Ponrathnam}, {Popolizio}, {Porter}, {Powell}, {Prajapati}, {Prasai},
  {Prasanna}, {Pratten}, {Prestegard}, {Principe}, {Prodi}, {Prokhorov},
  {Punturo}, {Puppo}, {P{\"u}rrer}, {Qi}, {Quetschke}, {Quinonez}, {Raab},
  {Raaijmakers}, {Radkins}, {Radulesco}, {Raffai}, {Rafferty}, {Raja}, {Rajan},
  {Rajbhandari}, {Rakhmanov}, {Ramirez}, {Ramos-Buades}, {Rana}, {Rao},
  {Rapagnani}, {Raymond}, {Razzano}, {Read}, {Regimbau}, {Rei}, {Reid},
  {Reitze}, {Rettegno}, {Ricci}, {Richardson}, {Richardson}, {Ricker},
  {Riemenschneider}, {Riles}, {Rizzo}, {Robertson}, {Robinet}, {Rocchi},
  {Rodriguez-Soto}, {Rolland}, {Rollins}, {Roma}, {Romanelli}, {Romano},
  {Romel}, {Romero-Shaw}, {Romie}, {Rose}, {Rose}, {Rose}, {Rosi{\'n}ska},
  {Rosofsky}, {Ross}, {Rowan}, {Rowlinson}, {Roy}, {Roy}, {Roy}, {Ruggi},
  {Rutins}, {Ryan}, {Sachdev}, {Sadecki}, {Sakellariadou}, {Salafia},
  {Salconi}, {Saleem}, {Salemi}, {Samajdar}, {Sanchez}, {Sanchez},
  {Sanchis-Gual}, {Sanders}, {Santiago}, {Santos}, {Sarin}, {Sassolas},
  {Sathyaprakash}, {Sauter}, {Savage}, {Savant}, {Sawant}, {Sayah}, {Schaetzl},
  {Schale}, {Scheel}, {Scheuer}, {Schmidt}, {Schnabel}, {Schofield},
  {Sch{\"o}nbeck}, {Schreiber}, {Schulte}, {Schutz}, {Schwarm}, {Schwartz},
  {Scott}, {Scott}, {Seidel}, {Sellers}, {Sengupta}, {Sennett}, {Sentenac},
  {Sequino}, {Sergeev}, {Setyawati}, {Shaddock}, {Shaffer}, {Sharifi},
  {Shahriar}, {Sharma}, {Sharma}, {Shawhan}, {Shen}, {Shikauchi}, {Shink},
  {Shoemaker}, {Shoemaker}, {Shukla}, {ShyamSundar}, {Siellez}, {Sieniawska},
  {Sigg}, {Singer}, {Singh}, {Singh}, {Singha}, {Singhal}, {Sintes}, {Sipala},
  {Skliris}, {Slagmolen}, {Slaven-Blair}, {Smetana}, {Smith}, {Smith},
  {Somala}, {Son}, {Soni}, {Sorazu}, {Sordini}, {Sorrentino}, {Souradeep},
  {Sowell}, {Spencer}, {Spera}, {Srivastava}, {Srivastava}, {Staats},
  {Stachie}, {Standke}, {Steer}, {Steinke}, {Steinlechner}, {Steinlechner},
  {Steinmeyer}, {Stevenson}, {Stocks}, {Stops}, {Stover}, {Strain}, {Stratta},
  {Strunk}, {Sturani}, {Stuver}, {Sudhagar}, {Sudhir}, {Summerscales}, {Sun},
  {Sunil}, {Sur}, {Suresh}, {Sutton}, {Swinkels}, {Szczepa{\'n}czyk}, {Tacca},
  {Tait}, {Talbot}, {Tanasijczuk}, {Tanner}, {Tao}, {T{\'a}pai}, {Tapia},
  {Tapia San Martin}, {Tasson}, {Taylor}, {Tenorio}, {Terkowski},
  {Thirugnanasambandam}, {Thomas}, {Thomas}, {Thompson}, {Thondapu}, {Thorne},
  {Thrane}, {Tinsman}, {Saravanan}, {Tiwari}, {Tiwari}, {Tiwari}, {Toland},
  {Tonelli}, {Tornasi}, {Torres-Forn{\'e}}, {Torrie}, {Tosta e Melo},
  {T{\"o}yr{\"a}}, {Travasso}, {Traylor}, {Tringali}, {Tripathee}, {Trovato},
  {Trudeau}, {Tsang}, {Tse}, {Tso}, {Tsukada}, {Tsuna}, {Tsutsui}, {Turconi},
  {Ubhi}, {Udall}, {Ueno}, {Ugolini}, {Unnikrishnan}, {Urban}, {Usman},
  {Utina}, {Vahlbruch}, {Vajente}, {Valdes}, {Valentini}, {van Bakel}, {van
  Beuzekom}, {van den Brand}, {Van Den Broeck}, {Vander-Hyde}, {van der
  Schaaf}, {Van Heijningen}, {van Veggel}, {Vardaro}, {Varma}, {Vass},
  {Vas{\'u}th}, {Vecchio}, {Vedovato}, {Veitch}, {Veitch}, {Venkateswara},
  {Venugopalan}, {Verkindt}, {Veske}, {Vetrano}, {Vicer{\'e}}, {Viets},
  {Vinciguerra}, {Vine}, {Vinet}, {Vitale}, {Vivanco}, {Vo}, {Vocca},
  {Vorvick}, {Vyatchanin}, {Wade}, {Wade}, {Wade}, {Walet}, {Walker},
  {Wallace}, {Wallace}, {Walsh}, {Wang}, {Wang}, {Wang}, {Ward}, {Warden},
  {Warner}, {Was}, {Watchi}, {Weaver}, {Wei}, {Weinert}, {Weinstein}, {Weiss},
  {Wellmann}, {Wen}, {We{\ss}els}, {Westhouse}, {Wette}, {Whelan}, {Whiting},
  {Whittle}, {Wilken}, {Williams}, {Willis}, {Willke}, {Winkler}, {Wipf},
  {Wittel}, {Woan}, {Woehler}, {Wofford}, {Wong}, {Wright}, {Wu}, {Wysocki},
  {Xiao}, {Yamamoto}, {Yang}, {Yang}, {Yang}, {Yap}, {Yazback}, {Yeeles}, {Yu},
  {Yu}, {Yuen}, {Zadro{\.Z}ny}, {Zadro{\.Z}ny}, {Zanolin}, {Zelenova},
  {Zendri}, {Zevin}, {Zhang}, {Zhang}, {Zhang}, {Zhao}, {Zhao}, {Zhou}, {Zhou},
  {Zhu}, {Zimmerman}, {Zucker}, {Zweizig}, {LIGO Scientific Collaboration}, \&
  {Virgo Collaboration}}]{GW190521}
{Abbott}, R., {Abbott}, T.~D., {Abraham}, S., {et~al.} 2020{\natexlab{a}},
  \href{http://dx.doi.org/10.1103/PhysRevLett.125.101102}{\JournalTitle{\prl},
  125, 101102}

\bibitem[{{Abbott} {et~al.}(2020{\natexlab{b}})}]{lvc2020cat}
{Abbott}, R., {et~al.} 2020{\natexlab{b}}, \JournalTitle{arXiv e-prints},
  arXiv:2010.14527

\bibitem[{{Adams} \& {Fatuzzo}(1996)}]{adams1996a}
{Adams}, F.~C., \& {Fatuzzo}, M. 1996,
  \href{http://dx.doi.org/10.1086/177318}{\JournalTitle{\apj}, 464, 256}

\bibitem[{{Adams} \& {Laughlin}(1996)}]{adams1996b}
{Adams}, F.~C., \& {Laughlin}, G. 1996,
  \href{http://dx.doi.org/10.1086/177717}{\JournalTitle{\apj}, 468, 586}

\bibitem[{{Anderson}(1994)}]{Anderson1994}
{Anderson}, K.~J. 1994, {Star Wars: The Jedi Academy Trilogy}, Vol.~1, {Jedi
  Search} (New York: Bantam Spectra)

\bibitem[{{Askar} {et~al.}(2017){Askar}, {Szkudlarek}, {Gondek-Rosi{\'n}ska},
  {Giersz}, \& {Bulik}}]{Askar2017}
{Askar}, A., {Szkudlarek}, M., {Gondek-Rosi{\'n}ska}, D., {Giersz}, M., \&
  {Bulik}, T. 2017,
  \href{http://dx.doi.org/10.1093/mnrasl/slw177}{\JournalTitle{\mnras}, 464,
  L36}

\bibitem[{{Astropy Collaboration} {et~al.}(2013){Astropy Collaboration},
  {Robitaille}, {Tollerud}, {Greenfield}, {Droettboom}, {Bray}, {Aldcroft},
  {Davis}, {Ginsburg}, {Price-Whelan}, {Kerzendorf}, {Conley}, {Crighton},
  {Barbary}, {Muna}, {Ferguson}, {Grollier}, {Parikh}, {Nair}, {Unther},
  {Deil}, {Woillez}, {Conseil}, {Kramer}, {Turner}, {Singer}, {Fox}, {Weaver},
  {Zabalza}, {Edwards}, {Azalee Bostroem}, {Burke}, {Casey}, {Crawford},
  {Dencheva}, {Ely}, {Jenness}, {Labrie}, {Lim}, {Pierfederici}, {Pontzen},
  {Ptak}, {Refsdal}, {Servillat}, \& {Streicher}}]{Astropy2013}
{Astropy Collaboration}, {Robitaille}, T.~P., {Tollerud}, E.~J., {et~al.} 2013,
  \href{http://dx.doi.org/10.1051/0004-6361/201322068}{\JournalTitle{\aap},
  558, A33}

\bibitem[{{Banerjee}(2017)}]{Banerjee2017}
{Banerjee}, S. 2017,
  \href{http://dx.doi.org/10.1093/mnras/stw3392}{\JournalTitle{\mnras}, 467,
  524}

\bibitem[{{Banerjee} \& {Kroupa}(2011)}]{banerjee2011}
{Banerjee}, S., \& {Kroupa}, P. 2011,
  \href{http://dx.doi.org/10.1088/2041-8205/741/1/L12}{\JournalTitle{\apjl},
  741, L12}

\bibitem[{{Bartko} {et~al.}(2010){Bartko}, {Martins}, {Trippe}, {Fritz},
  {Genzel}, {Ott}, {Eisenhauer}, {Gillessen}, {Paumard}, {Alexander},
  {Dodds-Eden}, {Gerhard}, {Levin}, {Mascetti}, {Nayakshin}, {Perets},
  {Perrin}, {Pfuhl}, {Reid}, {Rouan}, {Zilka}, \& {Sternberg}}]{bartko2010}
{Bartko}, H., {Martins}, F., {Trippe}, S., {et~al.} 2010,
  \href{http://dx.doi.org/10.1088/0004-637X/708/1/834}{\JournalTitle{\apj},
  708, 834}

\bibitem[{{Belczynski} {et~al.}(2016){Belczynski}, {Heger}, {Gladysz},
  {Ruiter}, {Woosley}, {Wiktorowicz}, {Chen}, {Bulik}, {O'Shaughnessy}, {Holz},
  {Fryer}, \& {Berti}}]{belczynski2016}
{Belczynski}, K., {Heger}, A., {Gladysz}, W., {et~al.} 2016,
  \href{http://dx.doi.org/10.1051/0004-6361/201628980}{\JournalTitle{\aap},
  594, A97}

\bibitem[{{Bennett} {et~al.}(2014){Bennett}, {Larson}, {Weiland}, \&
  {Hinshaw}}]{Bennett2014}
{Bennett}, C.~L., {Larson}, D., {Weiland}, J.~L., \& {Hinshaw}, G. 2014,
  \href{http://dx.doi.org/10.1088/0004-637X/794/2/135}{\JournalTitle{\apj},
  794, 135}

\bibitem[{{Berti} {et~al.}(2007){Berti}, {Cardoso}, {Gonzalez}, {Sperhake},
  {Hannam}, {Husa}, \& {Br{\"u}gmann}}]{Berti2007}
{Berti}, E., {Cardoso}, V., {Gonzalez}, J.~A., {et~al.} 2007,
  \href{http://dx.doi.org/10.1103/PhysRevD.76.064034}{\JournalTitle{\prd}, 76,
  064034}

\bibitem[{{Breen} \& {Heggie}(2013)}]{breene2013}
{Breen}, P.~G., \& {Heggie}, D.~C. 2013,
  \href{http://dx.doi.org/10.1093/mnras/stt628}{\JournalTitle{\mnras}, 432,
  2779}

\bibitem[{Chatterjee {et~al.}(2010)Chatterjee, Fregeau, Umbreit, \&
  Rasio}]{Chatterjee2010}
Chatterjee, S., Fregeau, J.~M., Umbreit, S., \& Rasio, F.~A. 2010,
  \href{http://adsabs.harvard.edu/cgi-bin/nph-data_query?bibcode=2010ApJ...719..915C&link_type=EJOURNAL
  papers3://publication/doi/10.1088/0004-637X/719/1/915}{\JournalTitle{\apj},
  719, 915}

\bibitem[{{Chatterjee} {et~al.}(2017){Chatterjee}, {Rodriguez}, \&
  {Rasio}}]{chatterjee2017}
{Chatterjee}, S., {Rodriguez}, C.~L., \& {Rasio}, F.~A. 2017,
  \href{http://dx.doi.org/10.3847/1538-4357/834/1/68}{\JournalTitle{\apj}, 834,
  68}

\bibitem[{Chatterjee {et~al.}(2013)Chatterjee, Umbreit, Fregeau, \&
  Rasio}]{Chatterjee2013}
Chatterjee, S., Umbreit, S., Fregeau, J.~M., \& Rasio, F.~A. 2013,
  \href{http://dx.doi.org/10.1093/mnras/sts464}{\JournalTitle{\mnras}, 429,
  2881}

\bibitem[{{Chernoff} \& {Weinberg}(1990)}]{ChernoffWeinberg1990}
{Chernoff}, D.~F., \& {Weinberg}, M.~D. 1990,
  \href{http://dx.doi.org/10.1086/168451}{\JournalTitle{\apj}, 351, 121}

\bibitem[{Clark(1975)}]{Clark1975}
Clark, G. 1975, \JournalTitle{\apj}, 199, L143

\bibitem[{{Dabringhausen} {et~al.}(2009){Dabringhausen}, {Kroupa}, \&
  {Baumgardt}}]{dabrin2009}
{Dabringhausen}, J., {Kroupa}, P., \& {Baumgardt}, H. 2009,
  \href{http://dx.doi.org/10.1111/j.1365-2966.2009.14425.x}{\JournalTitle{\mnras},
  394, 1529}

\bibitem[{{De Marchi} {et~al.}(2017){De Marchi}, {Panagia}, \&
  {Beccari}}]{demarchi2017}
{De Marchi}, G., {Panagia}, N., \& {Beccari}, G. 2017,
  \href{http://dx.doi.org/10.3847/1538-4357/aa85e9}{\JournalTitle{\apj}, 846,
  110}

\bibitem[{{De Marchi} {et~al.}(2007){De Marchi}, {Paresce}, \&
  {Pulone}}]{demarchi2007}
{De Marchi}, G., {Paresce}, F., \& {Pulone}, L. 2007,
  \href{http://dx.doi.org/10.1086/512856}{\JournalTitle{\apjl}, 656, L65}

\bibitem[{{Dehnen} \& {Binney}(1998)}]{Dehnen1998}
{Dehnen}, W., \& {Binney}, J. 1998,
  \href{http://dx.doi.org/10.1046/j.1365-8711.1998.01282.x}{\JournalTitle{\mnras},
  294, 429}

\bibitem[{{Di Carlo} {et~al.}(2020){Di Carlo}, {Mapelli}, {Bouffanais},
  {Giacobbo}, {Santoliquido}, {Bressan}, {Spera}, \& {Haardt}}]{dicarlo2020}
{Di Carlo}, U.~N., {Mapelli}, M., {Bouffanais}, Y., {et~al.} 2020,
  \href{http://dx.doi.org/10.1093/mnras/staa1997}{\JournalTitle{\mnras}, 497,
  1043}

\bibitem[{{Downing} {et~al.}(2010){Downing}, {Benacquista}, {Giersz}, \&
  {Spurzem}}]{downing2010}
{Downing}, J.~M.~B., {Benacquista}, M.~J., {Giersz}, M., \& {Spurzem}, R. 2010,
  \href{http://dx.doi.org/10.1111/j.1365-2966.2010.17040.x}{\JournalTitle{\mnras},
  407, 1946}

\bibitem[{{Downing} {et~al.}(2011){Downing}, {Benacquista}, {Giersz}, \&
  {Spurzem}}]{downing2011}
---. 2011,
  \href{http://dx.doi.org/10.1111/j.1365-2966.2011.19023.x}{\JournalTitle{\mnras},
  416, 133}

\bibitem[{{Duquennoy} \& {Mayor}(1991)}]{DuquennoyMayor1991}
{Duquennoy}, A., \& {Mayor}, M. 1991, \JournalTitle{\aap}, 500, 337

\bibitem[{{El-Badry} {et~al.}(2019){El-Badry}, {Quataert}, {Weisz}, {Choksi},
  \& {Boylan-Kolchin}}]{elbadry2019}
{El-Badry}, K., {Quataert}, E., {Weisz}, D.~R., {Choksi}, N., \&
  {Boylan-Kolchin}, M. 2019,
  \href{http://dx.doi.org/10.1093/mnras/sty3007}{\JournalTitle{\mnras}, 482,
  4528}

\bibitem[{{Fragione} \& {Banerjee}(2020)}]{fragban2020}
{Fragione}, G., \& {Banerjee}, S. 2020,
  \href{http://dx.doi.org/10.3847/2041-8213/abb671}{\JournalTitle{\apjl}, 901,
  L16}

\bibitem[{{Fragione} \& {Kocsis}(2018)}]{Fragione2018b}
{Fragione}, G., \& {Kocsis}, B. 2018,
  \href{http://dx.doi.org/10.1103/PhysRevLett.121.161103}{\JournalTitle{\prl},
  121, 161103}

\bibitem[{{Fragione} {et~al.}(2020{\natexlab{a}}){Fragione}, {Loeb}, \&
  {Rasio}}]{fraglr2020q}
{Fragione}, G., {Loeb}, A., \& {Rasio}, F.~A. 2020{\natexlab{a}},
  \href{http://dx.doi.org/10.3847/2041-8213/ab9093}{\JournalTitle{\apjl}, 895,
  L15}

\bibitem[{{Fragione} {et~al.}(2020{\natexlab{b}}){Fragione}, {Loeb}, \&
  {Rasio}}]{fraglr2020}
---. 2020{\natexlab{b}},
  \href{http://dx.doi.org/10.3847/2041-8213/abbc0a}{\JournalTitle{\apjl}, 902,
  L26}

\bibitem[{{Fregeau} {et~al.}(2003){Fregeau}, {G{\"u}rkan}, {Joshi}, \&
  {Rasio}}]{Fregeau2003}
{Fregeau}, J.~M., {G{\"u}rkan}, M.~A., {Joshi}, K.~J., \& {Rasio}, F.~A. 2003,
  \href{http://dx.doi.org/10.1086/376593}{\JournalTitle{\apj}, 593, 772}

\bibitem[{{Giersz} {et~al.}(2019){Giersz}, {Askar}, {Wang}, {Hypki}, {Leveque},
  \& {Spurzem}}]{giersz2019}
{Giersz}, M., {Askar}, A., {Wang}, L., {et~al.} 2019,
  \href{http://dx.doi.org/10.1093/mnras/stz1460}{\JournalTitle{\mnras}, 487,
  2412}

\bibitem[{{Giesler} {et~al.}(2018){Giesler}, {Clausen}, \& {Ott}}]{Giesler2018}
{Giesler}, M., {Clausen}, D., \& {Ott}, C.~D. 2018,
  \href{http://dx.doi.org/10.1093/mnras/sty659}{\JournalTitle{\mnras}, 477,
  1853}

\bibitem[{{Gonz{\'a}lez} {et~al.}(2020){Gonz{\'a}lez}, {Kremer}, {Chatterjee},
  {Fragione}, {Rodriguez}, {Weatherford}, {Ye}, \& {Rasio}}]{Gonzalez2020}
{Gonz{\'a}lez}, E., {Kremer}, K., {Chatterjee}, S., {et~al.} 2020,
  \JournalTitle{arXiv e-prints}, arXiv:2012.10497

\bibitem[{{Haghi} {et~al.}(2017){Haghi}, {Khalaj}, {Hasani Zonoozi}, \&
  {Kroupa}}]{haghi2017}
{Haghi}, H., {Khalaj}, P., {Hasani Zonoozi}, A., \& {Kroupa}, P. 2017,
  \href{http://dx.doi.org/10.3847/1538-4357/aa6719}{\JournalTitle{\apj}, 839,
  60}

\bibitem[{{Haghi} {et~al.}(2020){Haghi}, {Safaei}, {Zonoozi}, \&
  {Kroupa}}]{Haghi2020}
{Haghi}, H., {Safaei}, G., {Zonoozi}, A.~H., \& {Kroupa}, P. 2020,
  \href{http://dx.doi.org/10.3847/1538-4357/abbfb0}{\JournalTitle{\apj}, 904,
  43}

\bibitem[{{Harris} {et~al.}(2014){Harris}, {Morningstar}, {Gnedin},
  {O'Halloran}, {Blakeslee}, {Whitmore}, {C{\^o}t{\'e}}, {Geisler}, {Peng},
  {Bailin}, {Rothberg}, {Cockcroft}, \& {Barber DeGraaff}}]{Harris2014}
{Harris}, W.~E., {Morningstar}, W., {Gnedin}, O.~Y., {et~al.} 2014,
  \href{http://dx.doi.org/10.1088/0004-637X/797/2/128}{\JournalTitle{\apj},
  797, 128}

\bibitem[{{Heggie} \& {Hut}(2003)}]{HeggieHut2003}
{Heggie}, D., \& {Hut}, P. 2003, {The Gravitational Million-Body Problem: A
  Multidisciplinary Approach to Star Cluster Dynamics} (Cambridge: Cambridge
  Univ. Press)

\bibitem[{Heggie(1975)}]{Heggie1975}
Heggie, D.~C. 1975, \href{http://adsabs.harvard.edu/abs/1975MNRAS.173..729H
  papers3://publication/uuid/EEE9B361-0082-4772-8BB8-23C720F9704E}{\JournalTitle{\mnras},
  173, 729}

\bibitem[{H{\'{e}}non(1971{\natexlab{a}})}]{Henon1971a}
H{\'{e}}non, M. 1971{\natexlab{a}},
  \href{http://link.springer.com/article/10.1007/BF00649159
  papers3://publication/doi/10.1007/BF00649159}{\JournalTitle{Astrophysics and
  Space Science}, 13, 284}

\bibitem[{H{\'{e}}non(1971{\natexlab{b}})}]{Henon1971b}
---. 1971{\natexlab{b}},
  \href{http://dx.doi.org/10.1007/BF00649201}{\JournalTitle{Astrophysics and
  Space Science}, 14, 151}

\bibitem[{{Ivanova} {et~al.}(2008){Ivanova}, {Heinke}, {Rasio}, {Belczynski},
  \& {Fregeau}}]{Ivanova2008}
{Ivanova}, N., {Heinke}, C.~O., {Rasio}, F.~A., {Belczynski}, K., \& {Fregeau},
  J.~M. 2008,
  \href{http://dx.doi.org/10.1111/j.1365-2966.2008.13064.x}{\JournalTitle{\mnras},
  386, 553}

\bibitem[{Joshi {et~al.}(2001)Joshi, Nave, \& Rasio}]{Joshi2001}
Joshi, K.~J., Nave, C.~P., \& Rasio, F.~A. 2001,
  \href{http://adsabs.harvard.edu/cgi-bin/nph-data_query?bibcode=2001ApJ...550..691J&link_type=EJOURNAL
  papers3://publication/doi/10.1086/319771}{\JournalTitle{\apj}, 550, 691}

\bibitem[{Joshi {et~al.}(2000)Joshi, Rasio, Zwart, \&
  Portegies~Zwart}]{Joshi2000}
Joshi, K.~J., Rasio, F.~A., Zwart, S.~P., \& Portegies~Zwart, S. 2000,
  \href{http://dx.doi.org/10.1086/309350}{\JournalTitle{\apj}, 540, 969}

\bibitem[{{King}(1962)}]{King1962}
{King}, I. 1962, \href{http://dx.doi.org/10.1086/108756}{\JournalTitle{\aj},
  67, 471}

\bibitem[{{Kremer} {et~al.}(2018){Kremer}, {Chatterjee}, {Rodriguez}, \&
  {Rasio}}]{Kremer2018a}
{Kremer}, K., {Chatterjee}, S., {Rodriguez}, C.~L., \& {Rasio}, F.~A. 2018,
  \href{http://dx.doi.org/10.3847/1538-4357/aa99df}{\JournalTitle{\apj}, 852,
  29}

\bibitem[{{Kremer} {et~al.}(2020{\natexlab{a}}){Kremer}, {Ye}, {Chatterjee},
  {Rodriguez}, \& {Rasio}}]{Kremer2019d}
{Kremer}, K., {Ye}, C.~S., {Chatterjee}, S., {Rodriguez}, C.~L., \& {Rasio},
  F.~A. 2020{\natexlab{a}},
  \href{http://dx.doi.org/10.1017/S1743921319007269}{in Star Clusters: From the
  Milky Way to the Early Universe, ed. A.~{Bragaglia}, M.~{Davies}, A.~{Sills},
  \& E.~{Vesperini}, Vol. 351}, 357

\bibitem[{{Kremer} {et~al.}(2019){Kremer}, {Rodriguez}, {Amaro-Seoane},
  {Breivik}, {Chatterjee}, {Katz}, {Larson}, {Rasio}, {Samsing}, {Ye}, \&
  {Zevin}}]{Kremer2019b}
{Kremer}, K., {Rodriguez}, C.~L., {Amaro-Seoane}, P., {et~al.} 2019,
  \href{http://dx.doi.org/10.1103/PhysRevD.99.063003}{\JournalTitle{\prd}, 99,
  063003}

\bibitem[{{Kremer} {et~al.}(2020{\natexlab{b}}){Kremer}, {Ye}, {Rui},
  {Weatherford}, {Chatterjee}, {Fragione}, {Rodriguez}, {Spera}, \&
  {Rasio}}]{kremer2020ApJS}
{Kremer}, K., {Ye}, C.~S., {Rui}, N.~Z., {et~al.} 2020{\natexlab{b}},
  \href{http://dx.doi.org/10.3847/1538-4365/ab7919}{\JournalTitle{\apjs}, 247,
  48}

\bibitem[{{Kremer} {et~al.}(2020{\natexlab{c}}){Kremer}, {Spera}, {Becker},
  {Chatterjee}, {Di Carlo}, {Fragione}, {Rodriguez}, {Ye}, \&
  {Rasio}}]{krem2020coll}
{Kremer}, K., {Spera}, M., {Becker}, D., {et~al.} 2020{\natexlab{c}},
  \href{http://dx.doi.org/10.3847/1538-4357/abb945}{\JournalTitle{\apj}, 903,
  45}

\bibitem[{{Kroupa}(2001)}]{Kroupa2001}
{Kroupa}, P. 2001,
  \href{http://dx.doi.org/10.1046/j.1365-8711.2001.04022.x}{\JournalTitle{\mnras},
  322, 231}

\bibitem[{{Lada} \& {Lada}(2003)}]{LadaLada2003}
{Lada}, C.~J., \& {Lada}, E.~A. 2003,
  \href{http://dx.doi.org/10.1146/annurev.astro.41.011802.094844}{\JournalTitle{\araa},
  41, 57}

\bibitem[{{Larson}(1998)}]{larson1998}
{Larson}, R.~B. 1998,
  \href{http://dx.doi.org/10.1046/j.1365-8711.1998.02045.x}{\JournalTitle{\mnras},
  301, 569}

\bibitem[{{Lousto} {et~al.}(2010){Lousto}, {Campanelli}, {Zlochower}, \&
  {Nakano}}]{Lousto2010}
{Lousto}, C.~O., {Campanelli}, M., {Zlochower}, Y., \& {Nakano}, H. 2010,
  \href{http://dx.doi.org/10.1088/0264-9381/27/11/114006}{\JournalTitle{Classical
  and Quantum Gravity}, 27, 114006}

\bibitem[{Lyne {et~al.}(1987)Lyne, Brinklow, Middleditch, Kulkarni, Backer, \&
  Clifton}]{Lyne1987}
Lyne, A., Brinklow, A., Middleditch, J., {et~al.} 1987, \JournalTitle{\nat},
  328, 399

\bibitem[{{Mackey} {et~al.}(2007){Mackey}, {Wilkinson}, {Davies}, \&
  {Gilmore}}]{mackey2007}
{Mackey}, A.~D., {Wilkinson}, M.~I., {Davies}, M.~B., \& {Gilmore}, G.~F. 2007,
  \href{http://dx.doi.org/10.1111/j.1745-3933.2007.00330.x}{\JournalTitle{\mnras},
  379, L40}

\bibitem[{{Mackey} {et~al.}(2008){Mackey}, {Wilkinson}, {Davies}, \&
  {Gilmore}}]{mackey2008}
---. 2008,
  \href{http://dx.doi.org/10.1111/j.1365-2966.2008.13052.x}{\JournalTitle{\mnras},
  386, 65}

\bibitem[{{Marks} {et~al.}(2012){Marks}, {Kroupa}, {Dabringhausen}, \&
  {Pawlowski}}]{marks2012}
{Marks}, M., {Kroupa}, P., {Dabringhausen}, J., \& {Pawlowski}, M.~S. 2012,
  \href{http://dx.doi.org/10.1111/j.1365-2966.2012.20767.x}{\JournalTitle{\mnras},
  422, 2246}

\bibitem[{{Morscher} {et~al.}(2015){Morscher}, {Pattabiraman}, {Rodriguez},
  {Rasio}, \& {Umbreit}}]{morscher2015}
{Morscher}, M., {Pattabiraman}, B., {Rodriguez}, C., {Rasio}, F.~A., \&
  {Umbreit}, S. 2015,
  \href{http://dx.doi.org/10.1088/0004-637X/800/1/9}{\JournalTitle{\apj}, 800,
  9}

\bibitem[{Pattabiraman {et~al.}(2013)Pattabiraman, Umbreit, Liao, Choudhary,
  Kalogera, Memik, \& Rasio}]{Pattabiraman2013}
Pattabiraman, B., Umbreit, S., Liao, W.-k., {et~al.} 2013,
  \href{http://dx.doi.org/10.1088/0067-0049/204/2/15}{\JournalTitle{\apjs},
  204, 15}

\bibitem[{Rodriguez {et~al.}(2015)Rodriguez, Morscher, Pattabiraman,
  Chatterjee, Haster, \& Rasio}]{Rodriguez2015a}
Rodriguez, C.~L., Morscher, M., Pattabiraman, B., {et~al.} 2015,
  \href{http://dx.doi.org/10.1103/PhysRevLett.115.051101}{\JournalTitle{\prl},
  115, 051101}

\bibitem[{{Rodriguez} {et~al.}(2019){Rodriguez}, {Zevin}, {Amaro-Seoane},
  {Chatterjee}, {Kremer}, {Rasio}, \& {Ye}}]{Rodriguez2019}
{Rodriguez}, C.~L., {Zevin}, M., {Amaro-Seoane}, P., {et~al.} 2019,
  \href{http://dx.doi.org/10.1103/PhysRevD.100.043027}{\JournalTitle{\prd},
  100, 043027}

\bibitem[{{Rodriguez} {et~al.}(2020){Rodriguez}, {Kremer}, {Grudi{\'c}},
  {Hafen}, {Chatterjee}, {Fragione}, {Lamberts}, {Martinez}, {Rasio},
  {Weatherford}, \& {Ye}}]{Rodriguez2020}
{Rodriguez}, C.~L., {Kremer}, K., {Grudi{\'c}}, M.~Y., {et~al.} 2020,
  \href{http://dx.doi.org/10.3847/2041-8213/ab961d}{\JournalTitle{\apjl}, 896,
  L10}

\bibitem[{{Sigurdsson} \& {Phinney}(1995)}]{Sigurdsson1995}
{Sigurdsson}, S., \& {Phinney}, E.~S. 1995,
  \href{http://dx.doi.org/10.1086/192199}{\JournalTitle{\apjs}, 99, 609}

\bibitem[{{Tichy} \& {Marronetti}(2008)}]{Tichy2008}
{Tichy}, W., \& {Marronetti}, P. 2008,
  \href{http://dx.doi.org/10.1103/PhysRevD.78.081501}{\JournalTitle{\prd}, 78,
  081501}

\bibitem[{{Verbunt} {et~al.}(1984){Verbunt}, {van Paradijs}, \&
  {Elson}}]{Verbunt1984}
{Verbunt}, F., {van Paradijs}, J., \& {Elson}, R. 1984,
  \href{http://dx.doi.org/10.1093/mnras/210.4.899}{\JournalTitle{\mnras}, 210,
  899}

\bibitem[{{Wang}(2020)}]{wang2020}
{Wang}, L. 2020,
  \href{http://dx.doi.org/10.1093/mnras/stz3179}{\JournalTitle{\mnras}, 491,
  2413}

\bibitem[{{Ye} {et~al.}(2020){Ye}, {Fong}, {Kremer}, {Rodriguez}, {Chatterjee},
  {Fragione}, \& {Rasio}}]{ye2020}
{Ye}, C.~S., {Fong}, W.-f., {Kremer}, K., {et~al.} 2020,
  \href{http://dx.doi.org/10.3847/2041-8213/ab5dc5}{\JournalTitle{\apjl}, 888,
  L10}

\bibitem[{{Ye} {et~al.}(2019){Ye}, {Kremer}, {Chatterjee}, {Rodriguez}, \&
  {Rasio}}]{Ye2019}
{Ye}, C.~S., {Kremer}, K., {Chatterjee}, S., {Rodriguez}, C.~L., \& {Rasio},
  F.~A. 2019,
  \href{http://dx.doi.org/10.3847/1538-4357/ab1b21}{\JournalTitle{\apj}, 877,
  122}

\end{thebibliography}

\end{document}